\documentclass[fleqn,usenatbib]{mnras}

\usepackage[T1]{fontenc}

\DeclareRobustCommand{\VAN}[3]{#2}
\let\VANthebibliography\thebibliography
\def\thebibliography{\DeclareRobustCommand{\VAN}[3]{##3}\VANthebibliography}

\usepackage{graphicx}	
\usepackage{amsmath}	

\usepackage{newtxtext,newtxmath}

\newcommand{\Msun}{\,\rmn{M}_{\sun}}

\newcommand{\Gyr}{\,\rmn{Gyr}}
\newcommand{\Myr}{\,\rmn{Myr}}

\newcommand{\kpc}{\,\rmn{kpc}}
\newcommand{\Mpc}{\,\rmn{Mpc}}
\newcommand{\cMpc}{\,\rmn{cMpc}}
\newcommand{\MLsun}{\,(\rmn{M/L})_{\sun}}

\newcommand{\kms}{\,\rmn{km}\,\rmn{s}^{-1}}
\newcommand{\kappaco}{\kappa_\mathrm{co,rot}}
\newcommand{\sSFR}{\mathrm{sSFR}}

\newcommand{\SigAv}{\left< \Sigma \right>}
\newcommand{\HI}{{H~\textsc{i}}}
\newcommand{\ts}{\textsuperscript}
\newcommand{\subfind}{\textsc{subfind}}
\newcommand{\Sersic}{S\'{e}rsic}

\title[EAGLE morphology-density relation]{The galaxy morphology-density relation in the EAGLE simulation}

\author[J. Pfeffer et al.]{Joel Pfeffer,$^{1}$\thanks{E-mail: joel.pfeffer@icrar.org (JP)}
Mitchell K. Cavanagh,$^{1}$
Kenji Bekki,$^{1}$
Warrick J. Couch,$^{2}$
Michael J. Drinkwater,$^{3}$
\newauthor
Duncan A. Forbes,$^{2}$
B\"{a}rbel S. Koribalski$^{4,5}$
\\
$^{1}$International Centre for Radio Astronomy Research (ICRAR), M468, University of Western Australia, 35 Stirling Hwy, Crawley, WA 6009, Australia \\
$^{2}$Centre for Astrophysics \& Supercomputing, Swinburne University of Technology, Hawthorn VIC 3122, Australia \\
$^{3}$School of Mathematics and Physics, University of Queensland, Brisbane, Queensland 4072, Australia \\
$^{4}$Australia Telescope National Facility, CSIRO Astronomy and Space Science, P.O. Box 76, Epping, NSW 1710, Australia \\
$^{5}$School of Science, Western Sydney University, Locked Bag 1797, Penrith, NSW 2751, Australia 
}

\date{Accepted 2022 November 22. Received 2022 November 20; in original form 2022 May 29}

\pubyear{2022}

\begin{document}
\label{firstpage}
\pagerange{\pageref{firstpage}--\pageref{lastpage}}
\maketitle

\begin{abstract}
The optical morphology of galaxies is strongly related to galactic environment, with the fraction of early-type galaxies increasing with local galaxy density.
In this work we present the first analysis of the galaxy morphology-density relation in a cosmological hydrodynamical simulation. We use a convolutional neural network, trained on observed galaxies, to perform visual morphological classification of galaxies with stellar masses $M_\ast > 10^{10} \Msun$ in the EAGLE simulation into elliptical, lenticular and late-type (spiral/irregular) classes.
We find that EAGLE reproduces both the galaxy morphology-density and morphology-mass relations.
Using the simulations, we find three key processes that result in the observed morphology-density relation: (i) transformation of disc-dominated galaxies from late-type (spiral) to lenticular galaxies through gas stripping in high-density environments, (ii) formation of lenticular galaxies by merger-induced black hole feedback in low-density environments, and (iii) an increasing fraction of high-mass galaxies, which are more often elliptical galaxies, at higher galactic densities.
\end{abstract}




\section{Introduction}
\label{sec:intro}

The optical morphology of galaxies contains information about both their shape and recent star formation, and thus may offer clues into galaxy formation and evolution processes.
One of the key findings of such studies is that morphology depends on local galaxy density, with the fraction of elliptical (E) and lenticular (S0) galaxies increasing with local density and a corresponding decrease in the fraction of late-type (spiral) galaxies, i.e. the galaxy morphology-density relation \citep[e.g.][]{Oemler_74, Davis_and_Geller_76, Dressler_80a, Postman_and_Geller_84}.
Alternatively, the morphology-density relation may be viewed as a morphology-radius relation in clusters \citep*{Dressler_80a, Whitmore_and_Gilmore_91, Whitmore_Gilmore_and_Jones_93, Fasano_et_al_15}.

Given S0s have stellar discs it is often suggested S0s are formed as `stripped spirals', either by direct stripping of the interstellar medium or `starvation' following stripping of the circumgalactic medium \citep[e.g.][]{Spitzer_and_Baade_51, Gunn_and_Gott_72, Larson_et_al_80, Bekki_et_al_02, Bekki_and_Couch_11}.
Such a mechanism may explain the morphology-density relation for both S0 and spiral galaxies, as stripping mechanisms will naturally be environmentally dependent.
However, simple disc fading does not explain the larger bulge fractions of S0s compared to spiral galaxies \citep{Faber_and_Gallagher_76, Dressler_80a, Croom_et_al_21}.
This morphological transformation could potentially be explained instead by `galaxy harassment' \citep[i.e. multiple high-speed encounters between galaxies,][]{Moore_et_al_96, Moore_et_al_98, Moore_et_al_99}, though galaxies observed in intermediate-redshift ($z \sim 0.3$) clusters do not show evidence of distortion that might be expected from such a mechanism \citep{Couch_et_al_98}.
Alternatively, concentrated star formation in the late stages of gas stripping could result in bulge growth \citep{Bekki_and_Couch_11}.
A stripped-spiral origin (or other environmentally-dependent processes) also cannot explain the origin of field or isolated S0 galaxies, which account for $\sim 20$ per cent of the population, though may explain the increase above the field fraction \citep{Postman_and_Geller_84, van_der_Wel_et_al_10}.

Alternatively, S0 galaxies may form from galaxy mergers \citep[both major and minor, e.g.][]{Toth_and_Ostriker_92, Quinn_Hernquist_and_Fullagar_93, Walker_Mihos_and_Hernquist_96, Bekki_98, Diaz_et_al_18, Dolfi_et_al_21}.
Mergers may explain many properties of S0s, such as their structure, dynamics and scaling relations \citep{Eliche-Moral_et_al_12, Eliche-Moral_et_al_13, Borlaff_et_al_14, Querejeta_et_al_15, Tapia_et_al_17}.
In particular, galaxy mergers may provide an origin for field S0s, but mergers become less likely once a galaxy becomes a satellite in a cluster (\citealt{Ghigna_et_al_98}; see also Section~\ref{sec:enviro_processes}).

These considerations suggest S0 galaxies may not have a single origin.
Quiescent galaxies appear to be significantly flatter in denser environments \citep{van_der_Wel_et_al_10}, and recent work has established that S0s in clusters are more rotationally supported than those in the field \citep{Coccato_et_al_20, Deeley_et_al_20}.
These observations are consistent with a merger origin for galaxies in low-density environments and an increasing importance of stripped spirals at higher densities, in agreement with the findings of S0 galaxy formation in cosmological hydrodynamical simulations \citep{Deeley_et_al_21}.

The main scenario for the formation of elliptical galaxies is morphological transformation of galaxies through mergers \citep[e.g.][]{Toomre_and_Toomre_72, Barnes_92, Naab_and_Burkert_03, Khochfar_et_al_11}.
Thus, formation through mergers may explain why the elliptical fraction rises steeply at the highest galaxy number densities \citep{Dressler_80a}, where galaxy mergers would be expected to be most frequent.
The morphology-density relation also persists when only considering slow-rotating galaxies, which only account for one third of ellipticals \citep{Cappellari_et_al_11, DEugenio_et_al_13, Houghton_et_al_13, van_de_Sande_et_al_21}.
However, the centres of galaxy clusters appear to host a constant fraction of elliptical galaxies, regardless of density \citep{Whitmore_Gilmore_and_Jones_93}.
This results in a morphology-density relation for ellipticals that rises at higher densities for clusters with higher average density \citep{Houghton_15}.
\citet{van_der_Wel_et_al_10} suggest that these trends are in fact driven by mass, with massive galaxies, which are more often ellipticals, occupying the centres of galaxy groups/clusters.

Cosmological hydrodynamical simulations have now increased sufficiently in realism such that they can reproduce many properties of the evolving galaxy population \citep[see reviews by][]{Somerville_and_Dave_15, Vogelsberger_et_al_20}.
Beginning with initial conditions consistent with a $\Lambda$ Cold Dark Matter universe, such simulations model the growth and evolution of galaxies in a cosmological context context by including the effects of gravity, cooling of gas, formation of stars and energy feedback from star formation and supermassive black holes.
By modelling and following the evolution of galaxies through time, simulations provide important tools to interpret the wealth of information provided by observational surveys.

To date, most work on the morphology of galaxies in hydrodynamical simulations focuses on morphological indicators (shape, kinematic or other non-parametric diagnostics) rather than visual morphology \citep[e.g.][but see \citealt{Huertas-Company_et_al_19}]{Correa_et_al_17, Rodriguez-Gomez_et_al_19, Tacchella_et_al_19, Thob_et_al_19, Trayford_et_al_19, Bignone_et_al_20}.
This body of work has shown that modern hydrodynamical simulations produce galaxies with morphological diagnostics and correlations between morphology, colour/star-formation rate and galaxy mass boardly in agreement with observed galaxies.
Though physically motivated (and less time-demanding or subjective than visual classification), the various morphological indicators are generally not straightforward to compare with observed morphologies (Hubble type).

In this work we use simulations from the `Evolution and Assembly of GaLaxies and their Environments' (EAGLE) project \citep{Schaye_et_al_15, Crain_et_al_15}, combined with a convolutional neural network (CNN) trained on observed galaxies to perform morphological classification \citep{Cavanagh_et_al_22}, to investigate whether modern hydrodynamical simulations can reproduce the galaxy morphology-density relation and offer insights into its origin.
Using a CNN to perform `visual' classification of simulated galaxies means that morphological relations can be compared in a consistent manner with observed galaxies.
With this work we aim to test which physical processes drive the increase in lenticular and elliptical fractions in denser environments, and the origin of field early-type galaxy fractions.

This paper is structured as follows. In Section~\ref{sec:methods} we describe the EAGLE simulations and the CNN used to perform galaxy classification. Section~\ref{sec:results} presents the main results from this work, including the morphology-density relation in EAGLE and its origin in the simulations. Finally, in Section~\ref{sec:summary} we discuss the connection between the morphology-density relation and S0 galaxy formation and summarise the results of this work.

\section{Methods}
\label{sec:methods}

\subsection{EAGLE simulations}
\label{sec:EAGLE}

The EAGLE project is a campaign of cosmological hydrodynamical simulations of galaxy formation and evolution \citep[for full details, see][]{Schaye_et_al_15, Crain_et_al_15}.
The simulations assume cosmological parameters consistent with the \citet[namely, $\Omega_\Lambda = 0.693$, $\Omega_\mathrm{m} = 0.307$, $\Omega_\mathrm{b} = 0.04825$, $h = 0.6777$, $\omega_8 = 0.8288$]{Planck_2014_paperXVI_short}.
The simulations were performed with a highly-modified version of the $N$-body, TreePM, smoothed particle hydrodynamics code \textsc{Gadget 3} code \citep[last described by][]{Springel_05}.

Importantly, the simulations include routines for radiative cooling \citep*{Wiersma_Schaye_and_Smith_09}, star formation \citep{Schaye_and_Dalla_Vecchia_08}, stellar evolution \citep{Wiersma_et_al_09}, stellar feedback \citep{Dalla_Vecchia_and_Schaye_12}, black holes \citep{Rosas-Guevara_et_al_15} and active galactic nucleus (AGN) feedback \citep{Booth_and_Schaye_09}.
The stellar and AGN feedback prescriptions were calibrated so that the simulations reproduce the galaxy stellar mass function, galaxy sizes and black hole masses at $z \approx 0$ \citep{Crain_et_al_15}.

The simulations have since been shown to reproduce a broad range of galaxy population properties, including the evolution of galaxy masses and sizes \citep{Furlong_et_al_15, Furlong_et_al_17}, galaxy luminosities and colours \citep{Trayford_et_al_15}, cold gas properties \citep{Lagos_et_al_15, Marasco_et_al_16, Crain_et_al_17} and circumgalactic and intergalactic absorption system properties \citep{Rahmati_et_al_15, Oppenheimer_et_al_16, Turner_et_al_16}.
The simulations are therefore ideal for comparing with observed galaxy populations.

In this work, we analyse the EAGLE reference simulation \citep[Ref-L100N1504][]{Schaye_et_al_15} of a $100^3$~comoving~Mpc$^3$ (cMpc) periodic volume. The volume initially contains $1504^3$ gas and dark matter particles with initial masses of $1.81 \times 10^6 \Msun$ and $9.7 \times 10^6 \Msun$, respectively, and has a maximum gravitational softening length of $0.7 \kpc$ (for $z < 2.8$, scaling as $2.66$~comoving~kpc at earlier times). 
In Appendix~\ref{app:CNN_test} we also use the simulation of a smaller $50^3$~cMpc$^3$ volume at the same resolution (Ref-L050N752) to test the accuracy of the CNN.
Galaxies are identified in the simulation by first running a friends-of-friends algorithm \citep{Davis_et_al_85} to identify dark matter structures, and then using the \subfind\ algorithm \citep{Springel_et_al_01, Dolag_et_al_09} to identify bound structures (galaxies/subhaloes).
To follow the evolution of galaxies between successive snapshots, galaxy merger trees were constructed using the \textsc{D-Trees} algorithm \citep{Jiang_et_al_14, Qu_et_al_17}.
This data is available publically through the EAGLE database \citep{McAlpine_et_al_16}.

\subsection{Sample selection and analysis}
\label{sec:analysis}

We limit our analysis to galaxies with $M_\ast (r < 50 \kpc) > 10^{10} \Msun$, such that they are well resolved with $\gtrsim 10^4$ stellar particles.
Particles within galaxies are defined as those bound to the subhalo according to \subfind\ (i.e. unbound particles and those bound to other subhaloes are excluded).
This gives us a sample of 3,607 galaxies at $z=0$.

We calculate projected number densities ($\Sigma_N = N / (\pi d^2_N)$, where $d_N$ is the distance to the $N$th neighbour) for each galaxy by projecting galaxies along the $z$ axis of the simulation volume (using only galaxies with $M_\ast > 10^{10} \Msun$ for the neighbour counts).
To mimic observational radial velocity cuts \citep[i.e. $\pm 1000 \kms$ in][]{Baldry_et_al_06}, we use a maximum line-of-sight ($z$ axis) separation between galaxies of $1000 \kms / H_0 \approx 15 \Mpc$.
For each galaxy we calculate $\Sigma_3$ \citep[as in][]{Dressler_80a} and an average density $\left< \Sigma \right>$ \citep[the average of $\Sigma_4$ and $\Sigma_5$, as in][]{Baldry_et_al_06}.

Synthetic SDSS $g$-band images to be classified by the CNN were generated for each galaxy using the process described in \citet{Cavanagh_et_al_22}, which we briefly summarise here.
The images with $100 \times 100$ pixels of size $0.5 \kpc$ were generated for the galaxies using \textsc{SPHviewer} \citep{Benitez-Llambay_15}, with particles distributed according to their local smoothing length.
Galaxies were oriented face-on using the spin vector for star particles between $2.5$ and $30 \kpc$ from the centre of the galaxy.
Rest-frame SDSS $g$-band luminosities were calculated for each stellar particle using the \textsc{fsps} stellar population model \citep{Conroy_Gunn_and_White_09, Conroy_and_Gunn_10}.
Following \citet{Trayford_et_al_17}, young star particles ($< 100 \Myr$) and star-forming gas particles were resampled at higher resolution ($10^3 \Msun$) based on the star-formation rate of the particles, to ensure that (e.g.) the spiral arms of galaxies are adequately sampled.
The new particles adopt the same properties (position, metallicity, smoothing lengths) as the particle from which they were sampled, and new ages were randomly assigned between $0$ and $100 \Myr$.
For the image smoothing lengths, young ($<100 \Myr$) particles adopt the local SPH smoothing, while older stars use a local stellar smoothing length calculated from the nearest 64 neighbours.

\subsection{Galaxy classification}
\label{sec:CNN}

\begin{figure*}
  \centering
  \includegraphics[width=\textwidth]{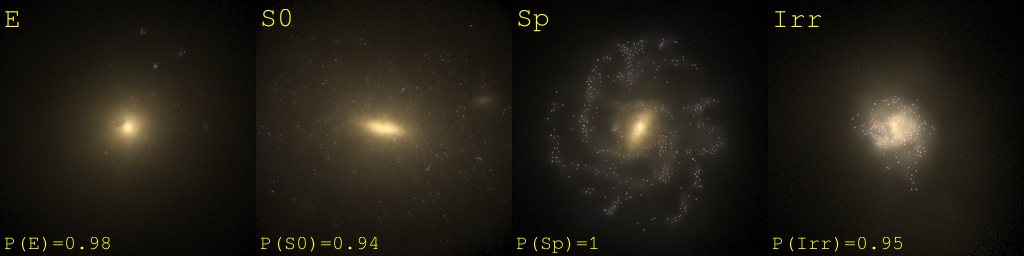}
  \caption{Examples of galaxies classified by the CNN as elliptical (left; $\mathrm{GalaxyID} = 19054212$), S0 (middle left; $\mathrm{GalaxyID} = 21109760$), spiral (middle right; $\mathrm{GalaxyID} = 16413979$) and irregular (right; $\mathrm{GalaxyID} = 9299365$). The images show face-on inclinations for each galaxy. The classification confidences are indicated in the bottom left of each panel. The mock three-colour ($gri$) images were generated for EAGLE galaxies by \citet{Trayford_et_al_17}, but note that CNN classifications use only $g$-band images.}
  \label{fig:examples}
\end{figure*}

Machine learning, and convolutional neural networks (CNNs) in particular, have been shown to be an extremely effective tool to perform image classification \citep*[see][for a review]{LeCun_Bengio_and_Hinton_15}.
In astronomy, CNNs have now been applied to a wide range of topics, such as classifying galaxy morphologies and bars, detecting mergers and identifying gravitational lenses \citep[e.g.][]{Dieleman_et_al_15, Abraham_et_al_18, Ackermann_et_al_18, Schaefer_et_al_18, Huertas-Company_et_al_19, Cavanagh_Bekki_and_Groves_21}.

To classify galaxy morphologies, we utilised a modified CNN based on the 4-class model (i.e. four classification categories) from \citet{Cavanagh_et_al_22}, originally developed in \citet{Cavanagh_Bekki_and_Groves_21}.
The CNN was trained on the \cite{Nair_and_Abraham_10} morphological catalogue and $g$-band imaging from the Sloan Digital Sky Survey DR7 \citep{York_et_al_00, Abazajian_et_al_09} to classify galaxies into E, S0, spirals and irregulars.
This catalogue contains 14,034 galaxies with redshifts $0.01 < z < 0.1$.
Though the difference between the training (real) and classification (simulated) images implies a domain shift\footnote{This case is reversed compared to the `reality gap' problem \citep{Tobin_et_al_17} that occurs when training on simulated images.}, the realism of the training set has more importance than that of the images being classified \citep[e.g. as found by][in the context of galaxy images]{Bottrell_et_al_19}. Therefore, the realism of the images being classified (e.g. lack of realistic fore-/backgrounds) is not expected to have a major impact on the results.

The core elements of a CNN are its convolutional layers, used to extract features using trainable filter weights, and pooling layers, used to downsample the input and preserve dominant features \citep{LeCun_Bengio_and_Hinton_15}.
The CNN used in this work is adapted from \citet{Cavanagh_et_al_22} and features four blocks of alternating convolutional and max pooling layers, followed by two fully-connected dense layers with 512 nodes each, for a total of 5.6 million trainable parameters.
Each convolutional layer utilises a $7\times7$ kernel with ReLU activation \citep{Glorot_et_al_11}.
We use ReLU activation for the dense layers, with softmax activation for the output layer.
We also use dropout layers after each pooling layer. Dropout is regularisation technique in which a set fraction of inputs are ignored; this helps mitigate overfitting and also improves model robustness \citep{Srivastava_et_al_14}.
The CNN was retrained from scratch with the new architecture.

All network parameters, including the learning rate for training, were further refined through extensive hyperparameter tuning using Optuna, an optimisation framework.
These changes further increased the accuracy of this CNN to 84 per cent, compared to 81 per cent in  \citet{Cavanagh_et_al_22}.
The network was trained and tested on an Nvidia RTX 3060 desktop GPU.
The classification of 35,082 EAGLE galaxies from \citet[with $M_\ast > 10^{10} \Msun$ and all snapshots in the redshift range $0 \leq z \leq 1$]{Cavanagh_et_al_22} took less than 2 minutes, though in this work we focus on the $z=0$ sample of 3,607 galaxies.
Our galaxy sample size is comparable to that from \citet{Huertas-Company_et_al_19}, who used a CNN to classify the morphologies of $z \approx 0$ galaxies from the IllustrisTNG simulation ($\sim$12,000 galaxies with $M_\ast > 10^{9.5} \Msun$).
We show examples of galaxies classified into each class (E, S0, Sp, Irr) in Fig.~\ref{fig:examples}. Each example galaxy has a classification confidence above 90 per cent (indicated in bottom left of each panel).

To demonstrate the accuracy of the CNN when applied to the simulated galaxy images, in Appendix~\ref{app:CNN_test} we compare the classifications of the CNN with human visual classifications for the EAGLE RefL050N752 simulation (i.e. the reference model for a smaller $50^3 \cMpc^3$ volume).
Two views of the galaxies (face- and edge-on) were used in the visual classifications to distinguish spheroidal and discy galaxies.
Overall, the CNN is 76-88 per cent accurate when considering early- (E+S0) and late-type (Sp+Irr) galaxies.
However, as discussed further in Appendix~\ref{app:CNN_test} and Section~\ref{sec:props}, flattened high-mass galaxies are often identified as ellipticals \citep[analogous to fast-rotating galaxies visually classified as ellipticals, e.g.][]{Cappellari_et_al_07, Emsellem_et_al_07}, while all low-mass early-type galaxies are classified by the CNN as S0s leading to underestimated elliptical fractions.

\section{Results}
\label{sec:results}

\subsection{Properties of morphological types}
\label{sec:props}

\begin{figure}
  \includegraphics[width=84mm]{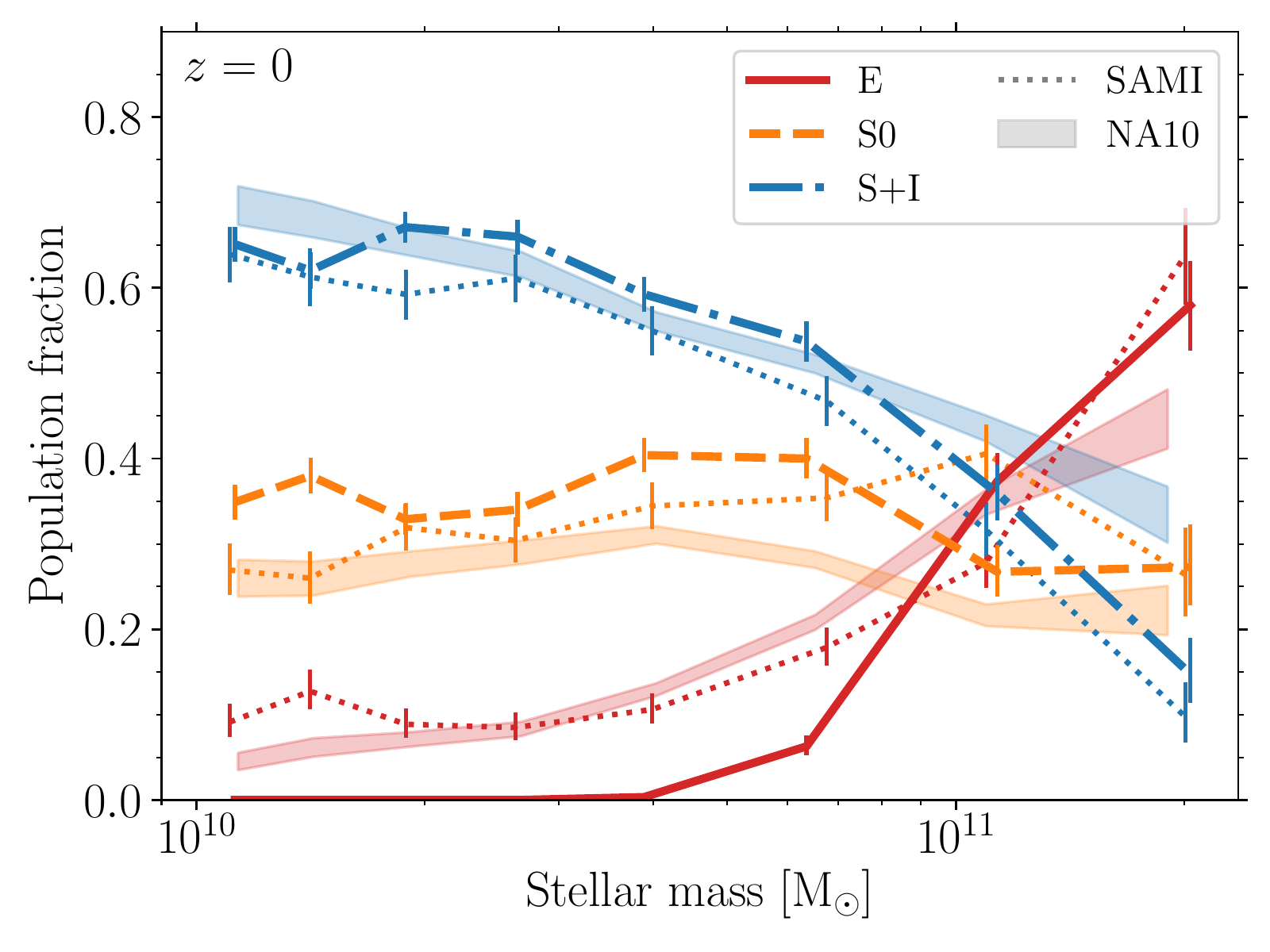}
  \caption{Dependence of morphology on galaxy stellar mass in the EAGLE simulation at $z=0$. Elliptical (E), lenticular (S0) and late-type (S+I) galaxies are shown as solid red, dashed and dash-dotted lines, respectively. Uncertainties in the figure are calculated using binomial statistics. For comparison, the shaded regions show the confidence intervals for the morphology-mass relations from \citet[using their sample for $z<0.03$]{Nair_and_Abraham_10}, while dotted lines show the SAMI survey \citep[$0.004 < z < 0.095$]{Bryant_et_al_15, Cortese_et_al_16}. Other than an absence of low-mass elliptical galaxies (discussed in Section~\ref{sec:props}), the EAGLE simulation shows good agreement with observed morphology fractions as a function of mass.}
  \label{fig:morph-mass}
\end{figure}

Before investigating the morphology-density relation (Section~\ref{sec:morph-dens}), we first verify the dependence of morphology on different galaxy properties.
In Fig.~\ref{fig:morph-mass} we compare the morphology-mass relations from EAGLE (3607 galaxies) with observed relations from \citet[2721 galaxies]{Nair_and_Abraham_10} and the Sydney-AAO Multi-object Integral field (SAMI) Galaxy Survey \citep[1823 galaxies, $0.004 < z < 0.095$]{Bryant_et_al_15, Cortese_et_al_16}.
The \citet{Nair_and_Abraham_10} catalogue has an apparent magnitude limit of SDSS $g < 16$~mag. Thus, we limit the catalogue to galaxies with redshifts $z<0.03$ to be reasonably complete for $M_\ast > 10^{10} \Msun$.
The predictions from the EAGLE simulation are generally in good agreement with the observed relations: the fraction of elliptical galaxies increases with stellar mass; the fraction of late-type galaxies decreases with stellar mass; while the fraction of lenticular galaxies is relatively constant with mass (decreasing only slightly at $M_\ast > 10^{11} \Msun$).
The strong increase in elliptical fraction at $M_\ast \gtrsim 10^{11}$ may be expected from the increasing contribution of mergers to galaxy stellar mass growth at higher masses \citep[e.g.][]{Robotham_et_al_14, Rodriguez-Gomez_et_al_16, Clauwens_et_al_18, Tacchella_et_al_19, Davison_et_al_20}.
At masses $>10^{11} \Msun$ the late-type fraction from EAGLE agrees better with the SAMI survey than the \citet{Nair_and_Abraham_10} sample.
The origin of this difference between the observational samples is not clear, but might simply be due to sample variance.

One issue with the classifications is the lack of EAGLE galaxies classified as elliptical at $M_\ast \lesssim 5 \times 10^{10} \Msun$, with a corresponding fraction of lenticular galaxies that is slightly too high compared to the observed fractions.
The likely origin of this issue is that the simulated galaxies are insufficiently concentrated due to numerical effects \citep[i.e. gravitational softening, the ISM cooling floor and/or two-body scattering due to more massive dark matter particles, see][]{Snyder_et_al_15, Bottrell_et_al_17, Ludlow_et_al_19, de_Graaff_et_al_21}.
Thus, otherwise spheroidal galaxies may be identified by the CNN as lenticular galaxies due to having density profiles closer to exponential discs.
We discuss this issue further in Appendix~\ref{app:CNN_test}.

\begin{figure*}
  \centering
  \includegraphics[width=0.9\textwidth]{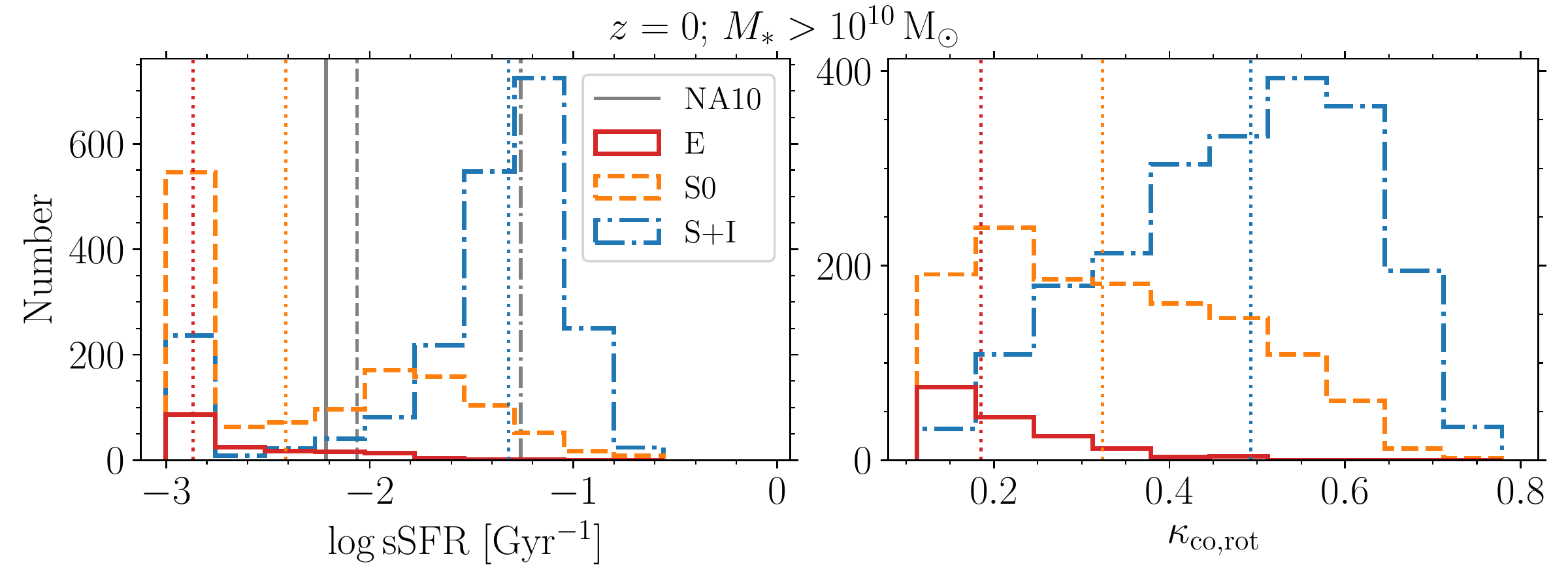}
  \caption{Distribution of specific star formation rate (sSFR) and $\kappaco$ (fraction of stellar kinetic energy in ordered co-rotation) for the different galaxy morphological types in the EAGLE simulation. Galaxies with low (or zero) specific star formation rates have been placed at $\sSFR = 10^{-3} \Gyr^{-1}$. The median values for each morphological types are shown as vertical dotted lines. For comparison, the median sSFRs from the \citet{Nair_and_Abraham_10} sample (as in Fig.~\ref{fig:morph-mass}) are shown as grey lines (with E, S0 and S+I galaxies shown as solid, dashed and dash-dotted lines, respectively). Morphologically-classified late-type (S+I) galaxies are star-forming and disc-dominated, while elliptical galaxies are quiescent and dispersion supported. S0 galaxies are typically intermediate between late-type and elliptical galaxies in both sSFR and $\kappaco$.}
  \label{fig:type-props}
\end{figure*}

Next, in Fig.~\ref{fig:type-props} we compare the distribution of specific star formation rates (sSFRs; using the instantaneous SFRs of gas particles bound to the subhalo and within $50\kpc$ of each galaxy) and the spin parameter $\kappaco$ (the fraction of stellar kinetic energy in ordered co-rotation).
In the left panel of Fig.~\ref{fig:type-props} we also show the median sSFR of different galaxy types from the \citet{Nair_and_Abraham_10} catalogue for reference \citep[where the SFRs are from][]{Brinchmann_et_al_04}.
EAGLE galaxies in the different morphological classes generally have properties expected for their given morphology:
Late-type galaxies (S+I) are typically star forming ($\sSFR \gtrsim 10^{-1.8} \Gyr^{-1}$, in good agreement with the observed galaxies) and disc dominated ($\kappaco \gtrsim 0.4$). Elliptical galaxies (E) are non-star forming ($\sSFR \sim 0$) and spheroidal ($\kappaco \sim 0.2$). Lenticular galaxies (S0) are intermediate between late-type and elliptical galaxies in both $\kappaco$ \citep[in qualitative agreement with angular momentum measurements,][]{Cortese_et_al_16, Falcon-Barroso_et_al_19} and sSFR, accounting for the majority of ``green valley'' ($10^{-2.8} \lesssim \sSFR / \Gyr^{-1} \lesssim 10^{-1.8}$) galaxies \citep[in agreement with][]{Bait_et_al_17}.
The sSFRs of E and S0 galaxies from \citet{Nair_and_Abraham_10} are somewhat higher than EAGLE galaxies. This may be due to the particular method for calculating sSFR \citep[using H$\alpha$ luminosity and 4000-\r{A} break,][]{Brinchmann_et_al_04}, as \citet{Bait_et_al_17} find lower sSFRs for quenched galaxies using spectral energy distribution fitting. However, qualitatively, the comparison shows that S0 galaxies in both observations and the EAGLE simulations have median sSFRs slightly higher than elliptical galaxies, but significantly lower than late-type galaxies.

As discussed above, with regard to the lack of lower-mass elliptical galaxies, many spheroidal galaxies ($\kappaco \lesssim 0.2$) are classified as lenticular galaxies.
A small fraction ($\sim 9$ per cent) of galaxies classified as late-type (S+I) also have $\sSFR = 0$. These galaxies are primarily from galaxy group/cluster regions (median $M_{200} = 10^{13.9} \Msun$, median $\Sigma_3 = 10^{1.2} \Mpc^{-2}$).
Some of these objects are galaxies classified as having irregular morphologies ($\approx$35 per cent of such galaxies).
Others may be due to misclassification by the CNN (recall the typical accuracy is 84 per cent).
From visual inspection, the majority of these cases would be classified by eye as S0 galaxies.

Elliptical galaxies feature a tail in $\kappaco$ toward rotationally-supported systems ($\kappaco \gtrsim 0.3$). This is somewhat similar to the distribution in stellar spin found for observed elliptical galaxies, i.e. morphologically-classified elliptical galaxies are a mix of `fast' and `slow' rotating galaxies \citep[and similarly some S0 galaxies have low stellar spin]{Cappellari_et_al_07, Emsellem_et_al_07, Emsellem_et_al_11}. 
The origin of slow rotators in the EAGLE simulations has been investigated in previous works \citep{Lagos_et_al_18b, Lagos_et_al_22}.
In this work we focus on the visual morphology-density relation, rather than the kinematic morphology-density relation \citep{Cappellari_et_al_11}, as most works to date have investigated the former.

Given the reasonable correspondence of the physical properties of different morphological types between the EAGLE simulation and observed galaxies, we now test whether EAGLE reproduces the observed galaxy morphology-density relation (Section~\ref{sec:morph-dens}) and what physical processes shape the relation in the simulation (Sections~\ref{sec:props-enviro} and \ref{sec:enviro_processes}).

\subsection{Morphology-density relation in EAGLE}
\label{sec:morph-dens}

\begin{figure*}
  \includegraphics[width=0.49\textwidth]{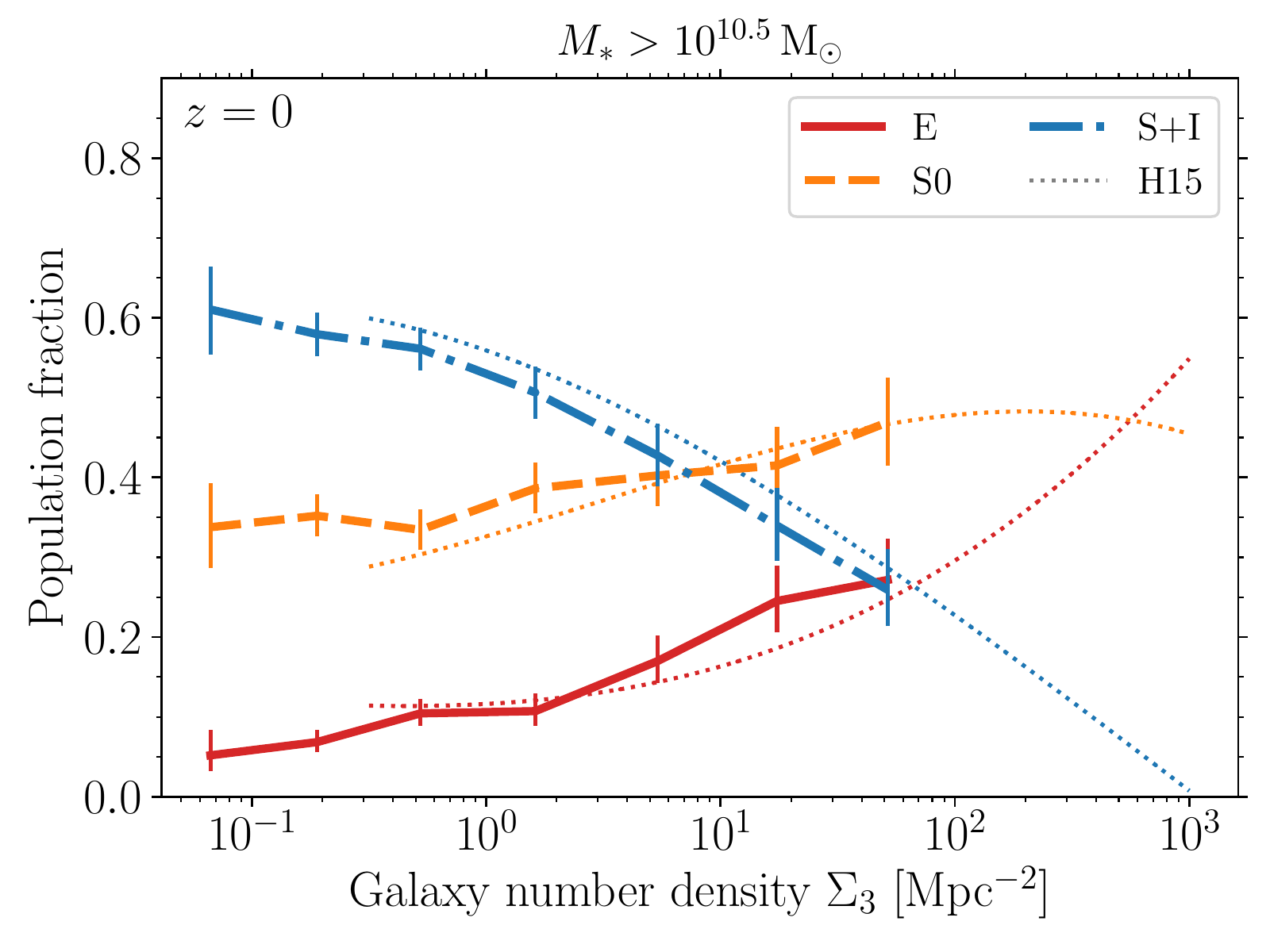}
  \includegraphics[width=0.49\textwidth]{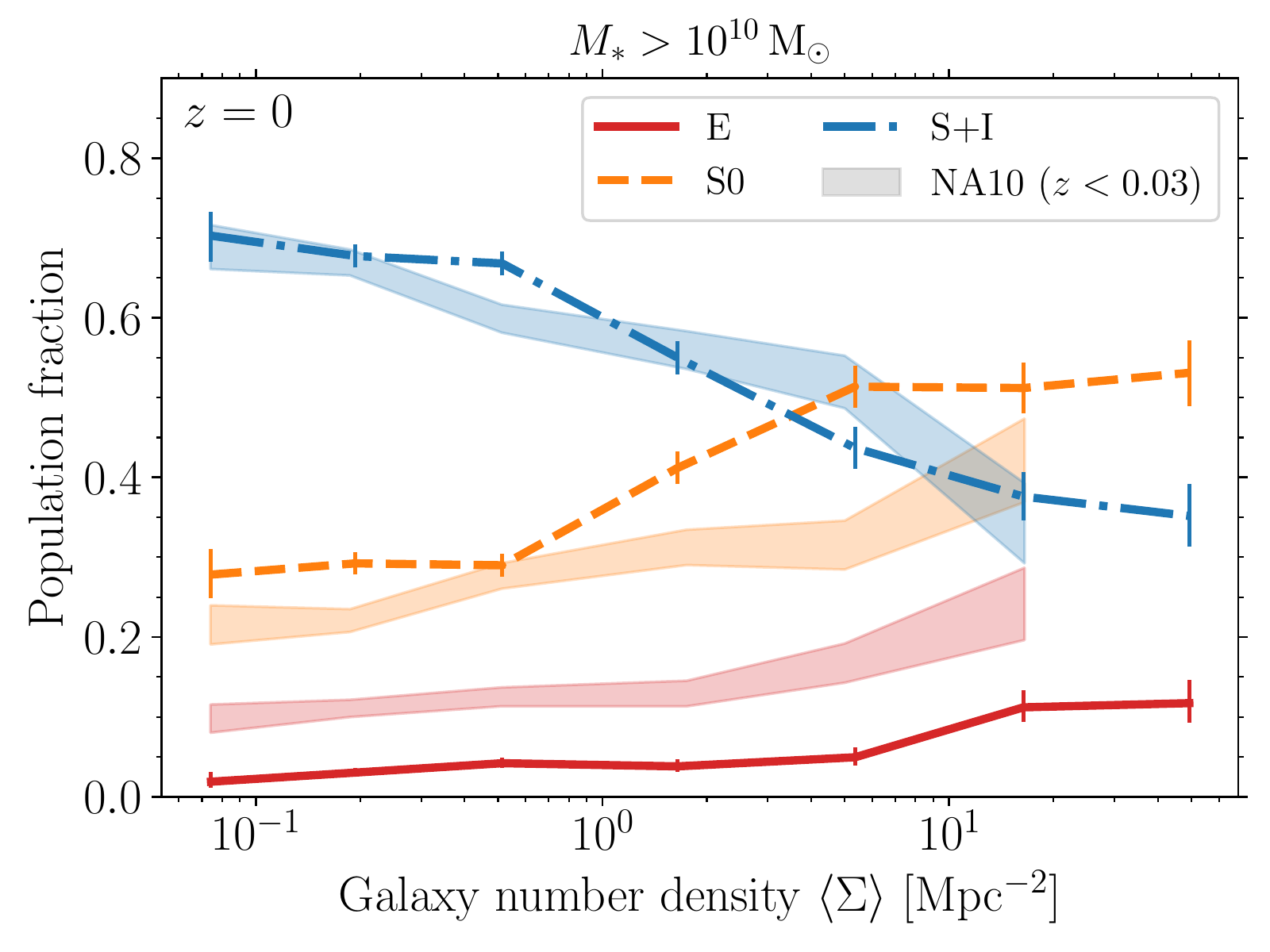}
  \caption{Morphology-density relations in the EAGLE simulation at $z=0$. Elliptical (E), lenticular (S0) and late-type (S+I) galaxies from EAGLE are shown as solid, dashed and dash-dotted lines, respectively. The left panel shows a comparison with the relations of observed galaxies from \citet[][an updated catalogue from \citealt{Dressler_80a}]{Houghton_15} as a function of $\Sigma_3$ (projected galaxy number density to 3\ts{rd} nearest neighbour). The right panel shows a comparison with the relations determined from the catalogue of \citet[for $z<0.03$]{Nair_and_Abraham_10} as a function of $\left< \Sigma \right>$ (average of $\Sigma_4$ and $\Sigma_5$). The title for each panel shows the mass limit applied in each comparison. We only show bins in $\Sigma$ where there are more than 50 galaxies, such that the results are not influenced by very low number statistics. Uncertainties in the figure are calculated using binomial statistics. The morphology-density relations from EAGLE agree very well with the \citet{Dressler_80a} and \citet{Houghton_15} relations for all morphology types. The late-type (S+I) relation from EAGLE also agrees well with the \citet{Nair_and_Abraham_10} catalogue, but the elliptical fraction is too low by $\approx 0.1$ due to issues with identification at low masses (see Section~\ref{sec:props} and Fig.~\ref{fig:morph-mass}).}
  \label{fig:morph-dens}
\end{figure*}

In Fig.~\ref{fig:morph-dens} we compare the morphology-density relation from the EAGLE simulation with observed relations from \citet[][a revised catalogue from \citealt{Dressler_80a}, with updated redshifts and densities]{Houghton_15} and \citet[restricted to $z < 0.03$]{Nair_and_Abraham_10}.
Given the dependence of morphology on galaxy mass (Fig.~\ref{fig:morph-mass}), we apply mass limits to the simulations to reasonably match observational limits.
The catalogue of \citet{Dressler_80a} has a galaxy luminosity limit $M_\mathrm{V} < -20.4$, or $M_\ast \gtrsim 10^{10.5} \Msun$ for a typical $M/L_\mathrm{V} \approx 2$-$3 \MLsun$.
We note that the galaxy sample from \citet{Nair_and_Abraham_10}, with median group halo masses of $\sim 10^{12} \Msun$, is more comparable to the EAGLE galaxy sample than the \citet{Dressler_80a} sample (55 rich galaxy clusters).

The left panel of Fig.~\ref{fig:morph-dens} shows the comparison of the EAGLE predictions with \citet[][showing the polynomial fits in their Table~1]{Houghton_15}, with the galaxy number density calculated using the distance to the third nearest galaxy, $\Sigma_3$.
Elliptical (E), lenticular (S0) and late-type (S+I) galaxy fractions from EAGLE are shown as solid, dashed and dash-dotted lines, respectively, while the fits from \citet{Houghton_15} are shown as dotted lines.
The volume of the EAGLE simulation ($100^3 \cMpc^3$) is not large enough to contain very massive ($\sim10^{15} \Msun$) galaxy clusters, and thus does not probe the highest densities of the observed relations.
Over the densities covered by the simulations, we find an extremely good match between the EAGLE and observed morphology-density relations.

The right panel of Fig.~\ref{fig:morph-dens} shows the comparison of EAGLE predictions with the catalogue of \citet{Nair_and_Abraham_10}, using the average galaxy number density (average of $\Sigma_4$ and $\Sigma_5$) from \citet{Baldry_et_al_06}.
The \citet{Nair_and_Abraham_10} results are shown as shaded regions (showing the uncertainty range calculated using binomial statistics).
As discussed in Section~\ref{sec:props}, there are too few elliptical galaxies identified by the CNN at low galaxy masses, resulting in a morphology-density relation for ellipticals that is too low by a fraction $\approx 0.1$.
However, the late-type (S+I) fraction from EAGLE is in good agreement with observed morphology-density relation.

Therefore, the EAGLE simulation reproduces the observed $z \approx 0$ galaxy morphology-density relation.
We stress here that these are genuine predictions from the EAGLE model.
The EAGLE galaxy formation model was calibrated with the $z\approx 0$ galaxy mass function, galaxy sizes and black hole masses \citep{Schaye_et_al_15}, not galaxy morphologies.

Given that the EAGLE model can reproduce the observed morphology-density relation, in the following sections we investigate what physical processes are important for shaping the relation in the simulation.

\subsection{Dependence of galaxy properties on environment}
\label{sec:props-enviro}

\begin{figure}
  \includegraphics[width=84mm]{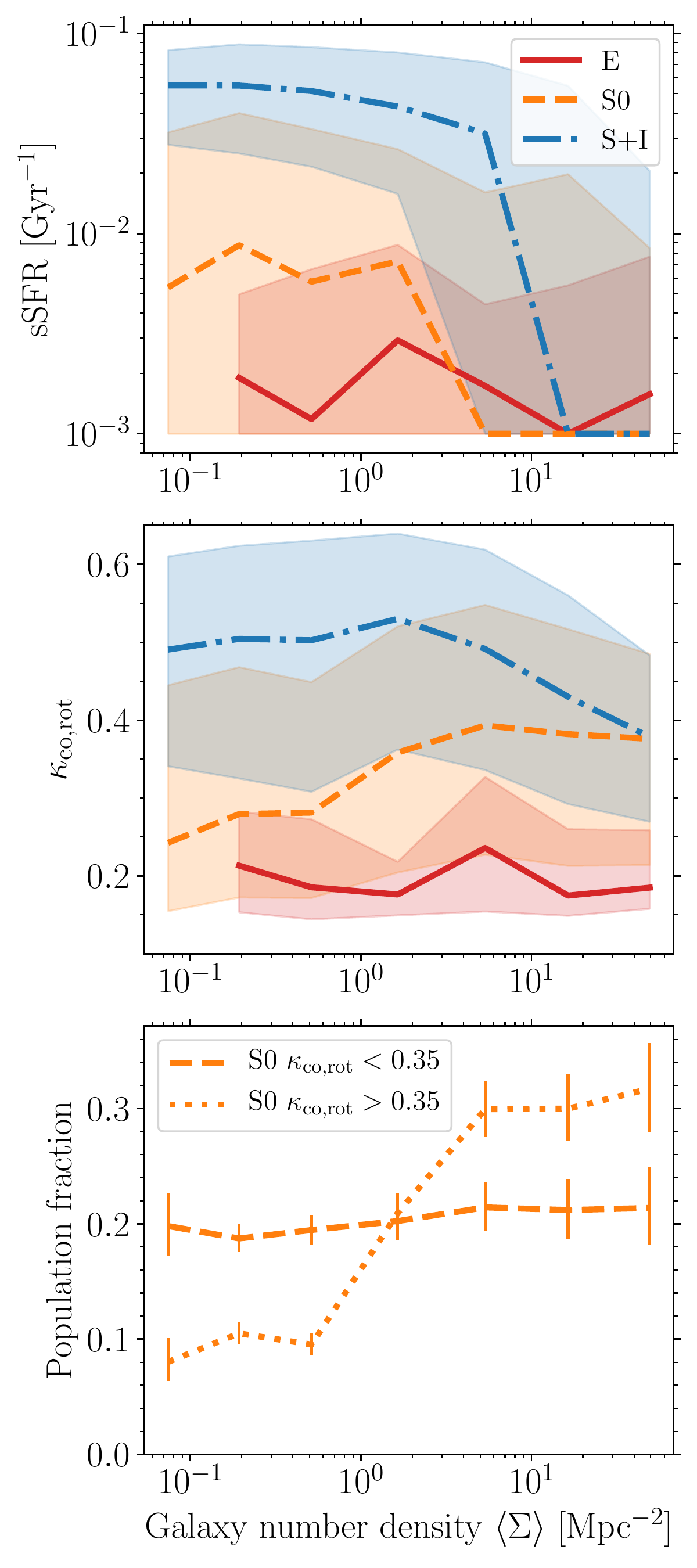}
  \caption{Properties for the different galaxy morphological types in the EAGLE simulation at $z=0$ as a function of galaxy number density (comparing the average density, as in the right panel of Fig.~\ref{fig:morph-dens}).  The panels show specific star formation rate (sSFR, upper panel), $\kappaco$ (fraction of stellar kinetic energy in ordered co-rotation, middle panel) and the fraction of the total galaxy population contributed by S0 galaxies with low (dashed line) and high (dotted line) rotation (lower panel). The thick lines show the median relation for each type, shaded regions show the 16\ts{th}-84\ts{th} percentiles. Solid red lines show elliptical galaxies (E), dashed orange lines show lenticular galaxies (S0) and dash-dotted blue lines show late-type galaxies (S+I). Galaxies with very low (typically zero) sSFRs have been placed at $\mathrm{sSFR} = 10^{-3} \Gyr^{-1}$. Only bins in density with at least 10 galaxies are shown.}
  \label{fig:props-dens}
\end{figure}

We compare the typical physical properties of galaxies as a function of density in Fig.~\ref{fig:props-dens}.
The figure shows how sSFR (upper panel) and $\kappaco$ (middle panel) correlate with density for each morphological type at $z=0$ (for galaxies with stellar masses $M_\ast > 10^{10} \Msun$ as in the right panel of Fig.~\ref{fig:morph-dens}).
As already expected from Fig.~\ref{fig:type-props}, S0 galaxies in low-density environments ($\left< \Sigma \right> \lesssim 3 \Mpc^{-2}$) are intermediate between late-type (S$+$I) and elliptical (E) galaxies in both sSFR and $\kappaco$. 
However, the figure shows that the properties of late-type and S0 galaxies correlate with environment: in the median, late-type galaxies show lower rotational support (lower $\kappaco$) at higher densities ($\left< \Sigma \right> \gtrsim 3 \Mpc^{-2}$), potentially because large rotationally-supported discs are impacted by dynamical processes in dense environments \citep[e.g. tidal stripping and harassment,][]{Moore_et_al_96, Moore_et_al_99}.
S0 galaxies show the inverse trend, becoming more rotationally supported (higher $\kappaco$) at higher densities (with similar median $\kappaco$ to S+I galaxies in the highest density bin).
\citet{Correa_et_al_17} similarly showed that the `red sequence' for EAGLE satellite galaxies has a much larger proportion of disc galaxies than central galaxies.
This trend for S0 galaxies is in agreement with the observational results by \citet{Coccato_et_al_20} and \citet{Deeley_et_al_20}, who found that S0 galaxies in groups and clusters have higher rotational support ($v / \sigma$) than field S0 galaxies.
Similarly, it is consistent with the kinematic morphology-density relation, where fast-rotating galaxies tend to become non-star forming (lenticular) at high densities \citep{Cappellari_et_al_11}.

The bottom panel of Fig.~\ref{fig:props-dens} shows the fraction of the total galaxy population that is contributed by S0 galaxies with high ($\kappaco > 0.35$) and low rotation ($\kappaco < 0.35$). 
Low rotation S0 galaxies contain a fixed fraction ($\approx0.2$) of the population at all densities, while the fraction of high rotation S0 galaxies increase with density.
Therefore, the increased S0 fraction in high-density environments in EAGLE appears to be due to the increase in quiescent disc galaxies, in agreement with the findings of previous studies \citep[e.g.][]{Postman_and_Geller_84, van_der_Wel_et_al_10, Deeley_et_al_20}.

At high densities, both S0 and late-type galaxies typically become completely non-star forming ($\mathrm{sSFR} = 0 \Gyr^{-1}$; at $\left< \Sigma \right> \gtrsim 5 \Mpc^{-2}$ for S0 galaxies, at $\left< \Sigma \right> \gtrsim 10 \Mpc^{-2}$ for late-type galaxies).
A similar transition to non-star forming galaxies at $\Sigma \gtrsim 10 \Mpc^{-2}$ for observed galaxies was found by \citet{Lewis_et_al_02}.
Possible causes are that the gas in the galaxies has been heated (e.g. via AGN feedback) such that it cannot cool to form stars, `strangulation' of star formation through prevention of gas accretion following cluster infall \citep{Larson_et_al_80} or that environmental processes such as tidal and ram pressure stripping \citep{Spitzer_and_Baade_51, Gunn_and_Gott_72} have completely removed the gas in these galaxies \citep[possibly combined with consumption of residual cold gas through star formation, e.g.][]{Larson_et_al_80, Bekki_et_al_02}. We discuss this further in Section~\ref{sec:high-density}.
Unlike S0 and late-type galaxies, the properties of elliptical galaxies (low sSFR and low $\kappaco$) are relatively constant across all environments.

\subsection{Physical drivers of the morphology-density relation}
\label{sec:enviro_processes}

\begin{figure*}
  \centering
  \includegraphics[width=\textwidth]{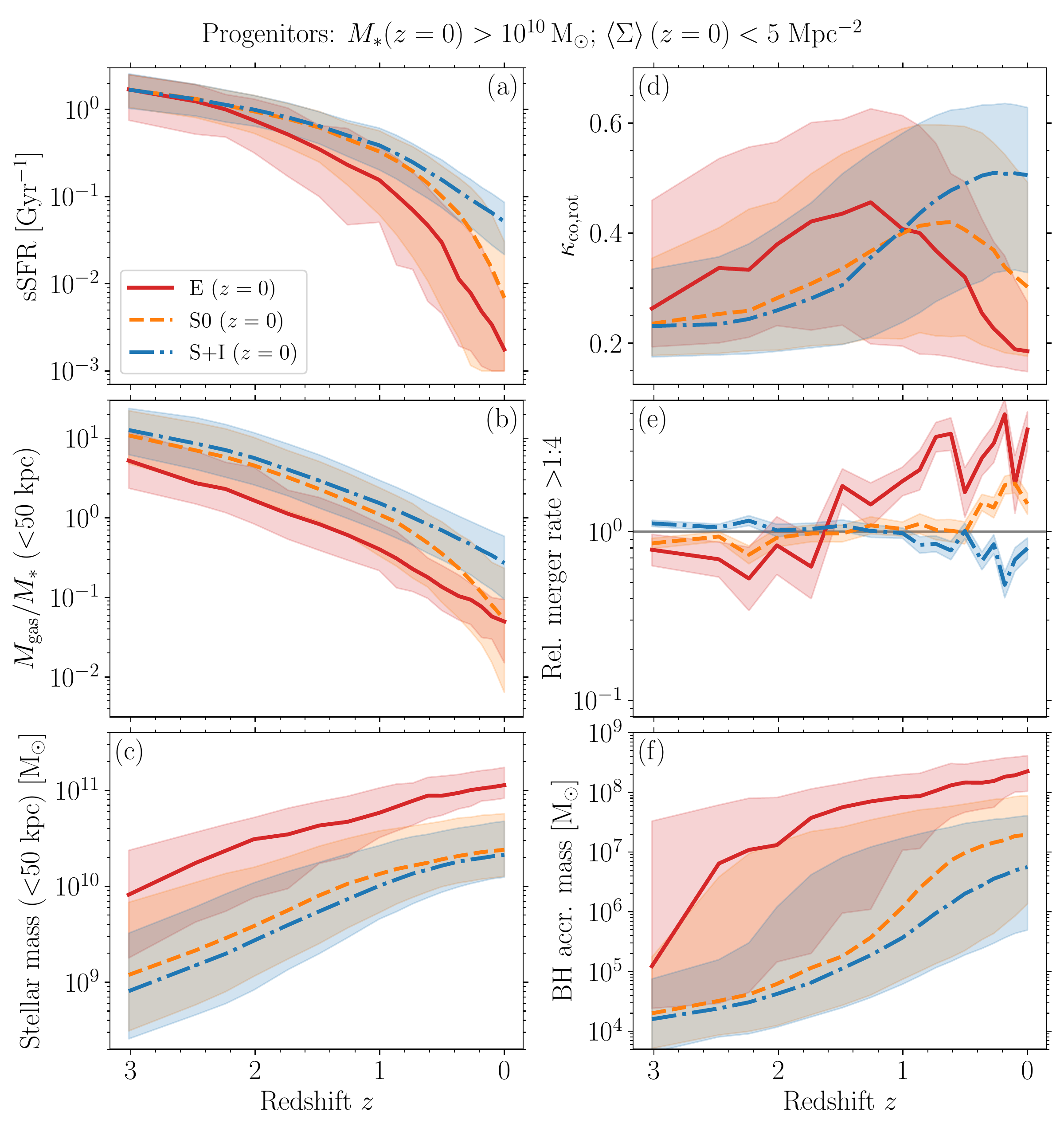}
  \caption{Redshift evolution of properties of main progenitors for galaxies residing in \textit{low} density environments ($\Sigma_3 < 5 \Mpc^{-2}$) at $z=0$ (high density environments are investigated in Fig.~\ref{fig:progs_highDens}). Galaxies are grouped according to their $z=0$ morphology, with solid red lines showing the median evolution of elliptical galaxies, dashed orange lines showing the median evolution of S0 galaxies and dash-dotted blue lines showing the median evolution of late-type galaxies. Shaded regions show the 16\ts{th}-84\ts{th} percentiles, except for panel (e) where the shaded regions show uncertainties in merger rates using binomial statistics. The subpanels of the figure show the evolution of: (a) specific star-formation rate; (b) gas-to-stellar mass ratio; (c) stellar mass within 50 kpc; (d) spin parameter $\kappaco$; (e) relative major merger rate (stellar mass ratio $>$1:4); (f) black hole accretion mass. The merger rates are calculated relative to the merger rate of all progenitors of galaxies with $M_\ast(z=0) > 10^{10} \Msun$ in the EAGLE volume.}
  \label{fig:progs_lowDens}
\end{figure*}

\begin{figure*}
  \centering
  \includegraphics[width=\textwidth]{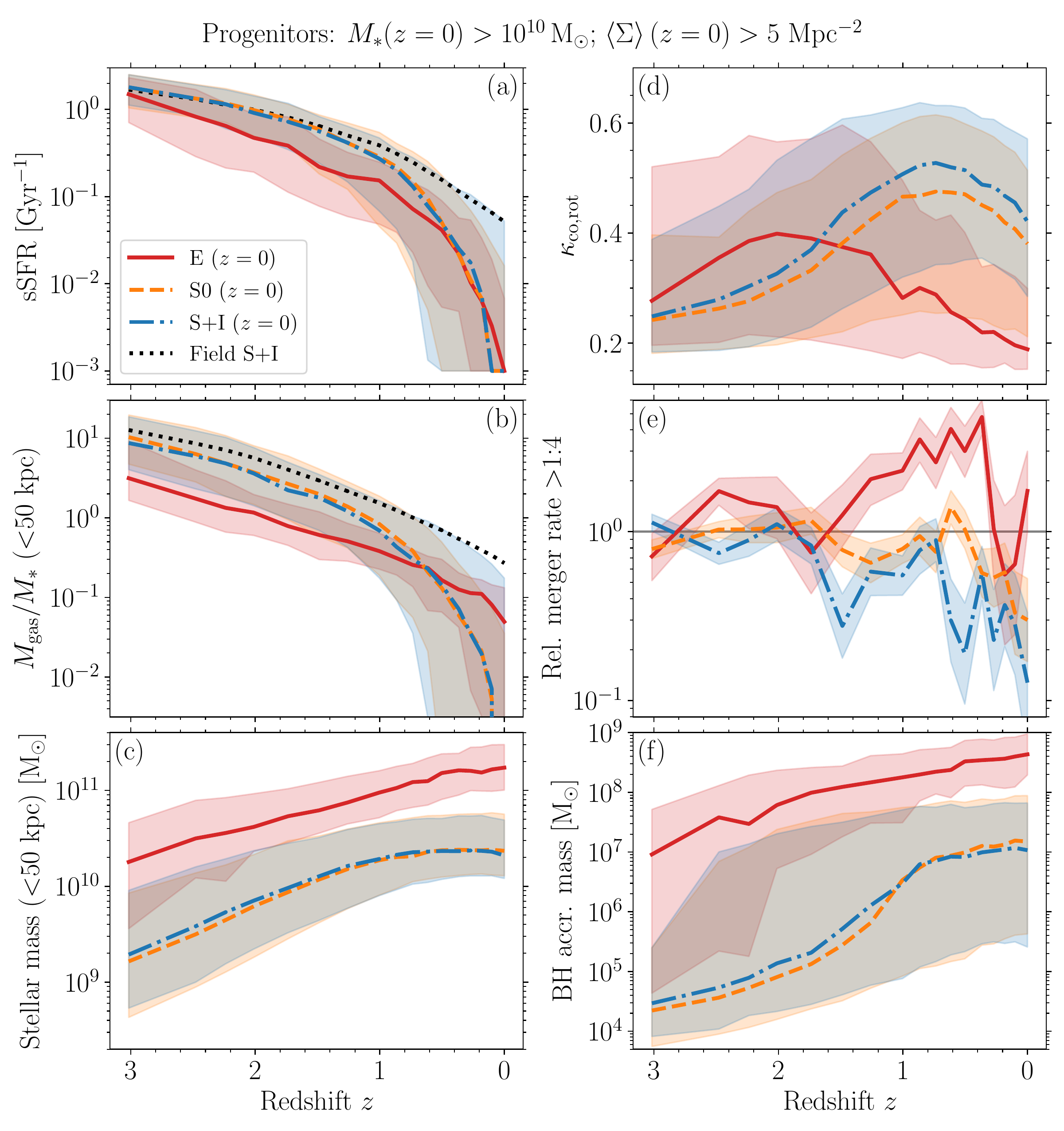}
  \caption{Redshift evolution of properties of main progenitors for galaxies residing in \textit{high} density environments ($\Sigma_3 > 5 \Mpc^{-2}$) at $z=0$. Panels and line styles are as in Fig.~\ref{fig:progs_lowDens}. For reference, the black dotted lines show the sSFR and gas-to-stellar mass ratio evolution of low-density S+I galaxies in Fig.~\ref{fig:progs_lowDens}.}
  \label{fig:progs_highDens}
\end{figure*}

To further investigate the formation and evolution processes affecting galaxies in different environments, in Figures~\ref{fig:progs_lowDens} and \ref{fig:progs_highDens} we compare the redshift evolution for the progenitors of different morphological types.
We divide the sample into `low' (Fig.~\ref{fig:progs_lowDens}) and `high' (Fig.~\ref{fig:progs_highDens}) density regions at $\SigAv = 5 \Mpc^{-2}$ (where above this density S0 fraction is highest, Fig~\ref{fig:morph-dens}, and sSFRs of S0 galaxies are lowest, Fig.~\ref{fig:props-dens}).
In each figure we compare the specific star formation rates (panel a), gas-to-stellar mass ratio (panel b), stellar mass (panel c), spin parameter $\kappaco$ (panel d), relative major merger rate (panel e) and black hole accretion mass (panel f).
The major merger (stellar mass ratio $>$1:4) rates are shown relative to the average merger rate of all progenitors of galaxies with $M_\ast(z=0) > 10^{10} \Msun$ in the EAGLE volume, in order to highlight the differences between each galaxy type. The merger times are measured at the snapshot where the galaxies have fully merged according to the \subfind\ algorithm, though we note the merging process may begin a number of snapshots prior to this time. To determine mass ratios, we calculate the ratio at the snapshot where the `satellite' galaxy was at its maximum stellar mass \citep[excluding snapshots when the main branch and satellite have `exchanged' mass temporarily, see][]{Qu_et_al_17}.
In the rest of the section we discuss low-density environments in Section~\ref{sec:low-density}, high-density environments in Section~\ref{sec:high-density} and focus on elliptical galaxies in Section~\ref{sec:ellipticals}.

\subsubsection{Low-density environments}
\label{sec:low-density}

In Fig.~\ref{fig:progs_lowDens} we compare the evolution of galaxies which are in low-density environments at $z=0$.
At $z \gtrsim 1$, the progenitors of S0 and late-type galaxies show a very similar evolution in both sSFR and $\kappaco$.
The galaxy types then diverge at $z \lesssim 1$, with late-type galaxies becoming more disc-dominated, and S0 galaxies becoming more dispersion supported and less star forming.
We argue this is due to a difference in galaxy merger rates: at $z < 1$ the progenitors of S0 galaxies experience a higher merger rate than late-type progenitors (by a factor $\sim 2$-$3$ at $z < 0.4$, well above the expected uncertainties in the merger rates from binomial statistics), which may explain the significantly lower $\kappaco$ for S0 galaxies.
This explanation is further supported by considering the progenitors of elliptical galaxies, which have the lowest median $\kappaco$ at $z<1$ and a merger rate nearly twice that of S0 progenitors over the same period.\footnote{We note that the transformation of the progenitors of massive elliptical galaxies from rotation to dispersion supported systems may driven by frequent minor mergers, rather than major mergers \citep[e.g.][]{Naab_Johansson_and_Ostriker_09, Khochfar_et_al_11}. However the relative merger rates in Fig.~\ref{fig:progs_lowDens} are quantitatively consistent when also considering minor mergers ($>$1:10 mass ratio), i.e. major merger rate is correlated with overall merger rate. We focus on major mergers here because of their connections between gas inflows and AGN feedback \citep[e.g.][]{Hernquist_89, Springel_DiMatteo_and_Hernquist_05, Davies_Pontzen_and_Crain_22}.}

Potentially, the difference in sSFRs between S0 and late-type progenitors at $z < 1$ may also be related to mergers, but modulated by AGN feedback.
Previous works have shown that star formation quenching of central galaxies in EAGLE is directly related to black hole growth and feedback \citep[e.g.][]{Bower_et_al_17, Correa_et_al_19, Davies_et_al_20}.
In the EAGLE model the energy released via AGN feedback is directly proportional to the mass accretion rate \citep{Booth_and_Schaye_09, Schaye_et_al_15}. 
Galaxies for which black holes have accreted more mass will have received more energy feedback from AGN (and thus are expected to have lower SFRs) than galaxies where black holes have accreted less mass \citep[e.g.][]{DiMatteo_Springel_and_Hernquist_05, Sijacki_et_al_07, Sijacki_et_al_15, Bower_et_al_17}.

We investigate the mass accreted by black holes in panel (f) of Fig.~\ref{fig:progs_lowDens}.
As expected \citep[given the known relationship between black hole mass and bulge mass/velocity dispersion, e.g.][]{Kormendy_and_Richstone_95, Magorrian_et_al_98, Ferrarese_and_Merritt_00, Savorgnan_et_al_16}, despite having similar median stellar masses (panel c), S0 galaxies have median black hole accretion masses over three times as large as late-type galaxies at $z=0$ ($\approx 2 \times 10^7 \Msun$ compared to $\approx 6 \times 10^6 \Msun$, respectively).
The median evolution of black hole accretion mass between late-type and S0 galaxies is largely similar at $z>1.2$.
The black holes in S0 galaxies then typically undergo a `rapid growth phase' \citep{Bower_et_al_17} at $z \approx 1.2$-$0.6$, before returning to a low accretion-rate phase ($z \lesssim 0.6$).
This rapid growth phase is consistent with the time at which the sSFRs of S0 and late-type galaxies begin to diverge (panel a).
In contrast, black holes in late-type galaxies typically have not undergone this rapid growth phase, leading to lower black hole accretion masses and higher SFRs.
We also investigated the gas fractions of the progenitor mergers \citep[following][]{Lagos_et_al_18a}, but found little difference in the fraction of gas-rich (`wet') mergers between S0 and late-type galaxies.

Previous works have shown that major galaxy mergers can disrupt the rotational motion of gas and drive strong inflows \citep[e.g.][]{Hernquist_89, Barnes_and_Hernquist_96, Mihos_and_Hernquist_96}, which may then drive rapid black hole growth and strong AGN feedback \citep{DiMatteo_Springel_and_Hernquist_05, Springel_DiMatteo_and_Hernquist_05, Hani_et_al_18, McAlpine_et_al_20, Davies_Pontzen_and_Crain_22}.
In turn, AGN feedback prevents the replenishment of the interstellar medium through heating and expulsion of the circumgalactic medium \citep{Davies_et_al_19, Davies_et_al_20, Oppenheimer_et_al_20}.
Mergers alone are unlikely to affect SFRs long term as, without some process to heat or expel gas, gaseous discs may survive or re-form after the merger \citep{Barnes_and_Hernquist_96, Springel_and_Hernquist_05, Robertson_et_al_06, Governato_et_al_07, Hopkins_et_al_09} and the merger remnant may remain star forming \citep{DiMatteo_Springel_and_Hernquist_05, Springel_DiMatteo_and_Hernquist_05, Davies_Pontzen_and_Crain_22}.
Similarly, AGN feedback alone is unlikely to significantly alter the kinematic properties of the stars in the galaxy (i.e. $\kappaco$ in panel d).
Thus, we arrive at a coherent picture in which the higher merger rates of the progenitors of field S0 and elliptical galaxies results in both their lower rotational support \citep{Cortese_et_al_16, Falcon-Barroso_et_al_19} and lower SFRs \citep{Bait_et_al_17} compared to late-type galaxies.
We note that the difference in merger rates between S0 and late-type galaxies is most different at $z < 0.4$, while the sSFRs diverge somewhat earlier at $z \sim 0.8$-$0.6$. This may be evidence of a `delay' \citep[see][]{Davies_Pontzen_and_Crain_22} between initial galaxy interactions (that triggers the black hole growth and AGN feedback) and the final coalescence of the galaxies (the snapshot at which the merger is identified in the galaxy merger tree).

\subsubsection{High-density environments}
\label{sec:high-density}

\begin{figure}
  \centering
  \includegraphics[width=84mm]{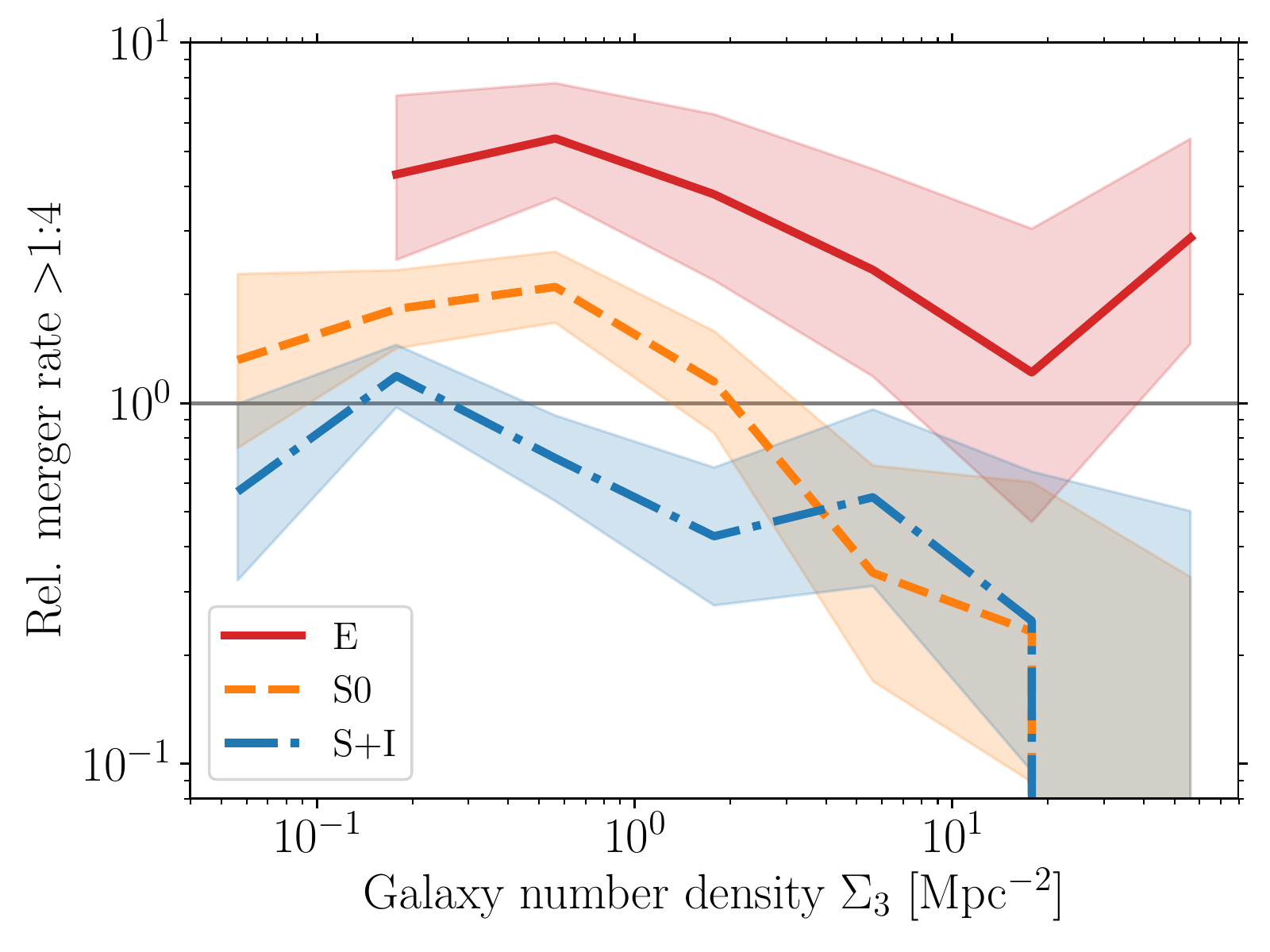}
  \caption{Relative major merger rates (stellar mass ratio $>1:4$) of galaxies at $z=0$ as a function of environment (galaxy number density) and galaxy morphological type. Galaxies are binned in density in an identical manner to Fig.~\ref{fig:morph-dens}, and only show density bins with a least 20 galaxies of a given morphological type. Shaded regions show the uncertainties on the merger rates from binomial statistics. The solid horizontal line shows the unity line for reference. As expected from $N$-body simulations \citep{Ghigna_et_al_98}, mergers become less likely for galaxies in higher density environments.}
  \label{fig:enviro_merger_rates}
\end{figure}

\begin{figure*}
  \centering
  \includegraphics[width=\textwidth]{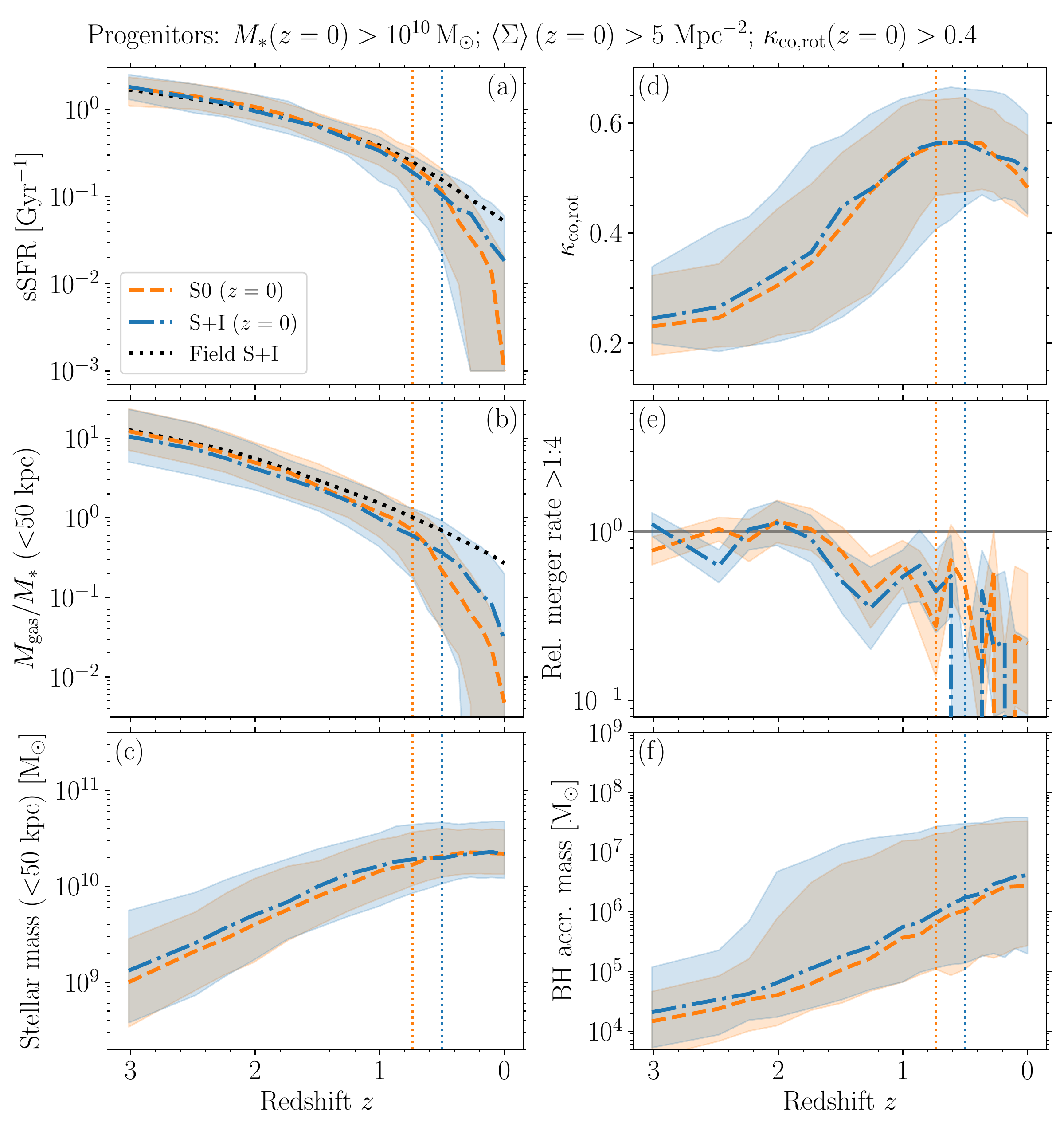}
  \caption{Redshift evolution of properties of main progenitors for galaxies residing in \textit{high} density environments ($\Sigma_3 > 5 \Mpc^{-2}$) and with $\kappaco > 0.4$ (disc-dominated) at $z=0$. Panels and line styles are as in Fig.~\ref{fig:progs_lowDens}. For reference, the black dotted lines show the sSFR and gas-to-stellar mass ratio evolution of low-density S+I galaxies in Fig.~\ref{fig:progs_lowDens}. Satellite S0 and S+I galaxies have median infall redshifts of $0.74$ and $0.5$, respectively, indicated by vertical dotted lines in each panel.}
  \label{fig:progs_highDens_highKappa}
\end{figure*}

\begin{figure*}
  \centering
  \includegraphics[width=\textwidth]{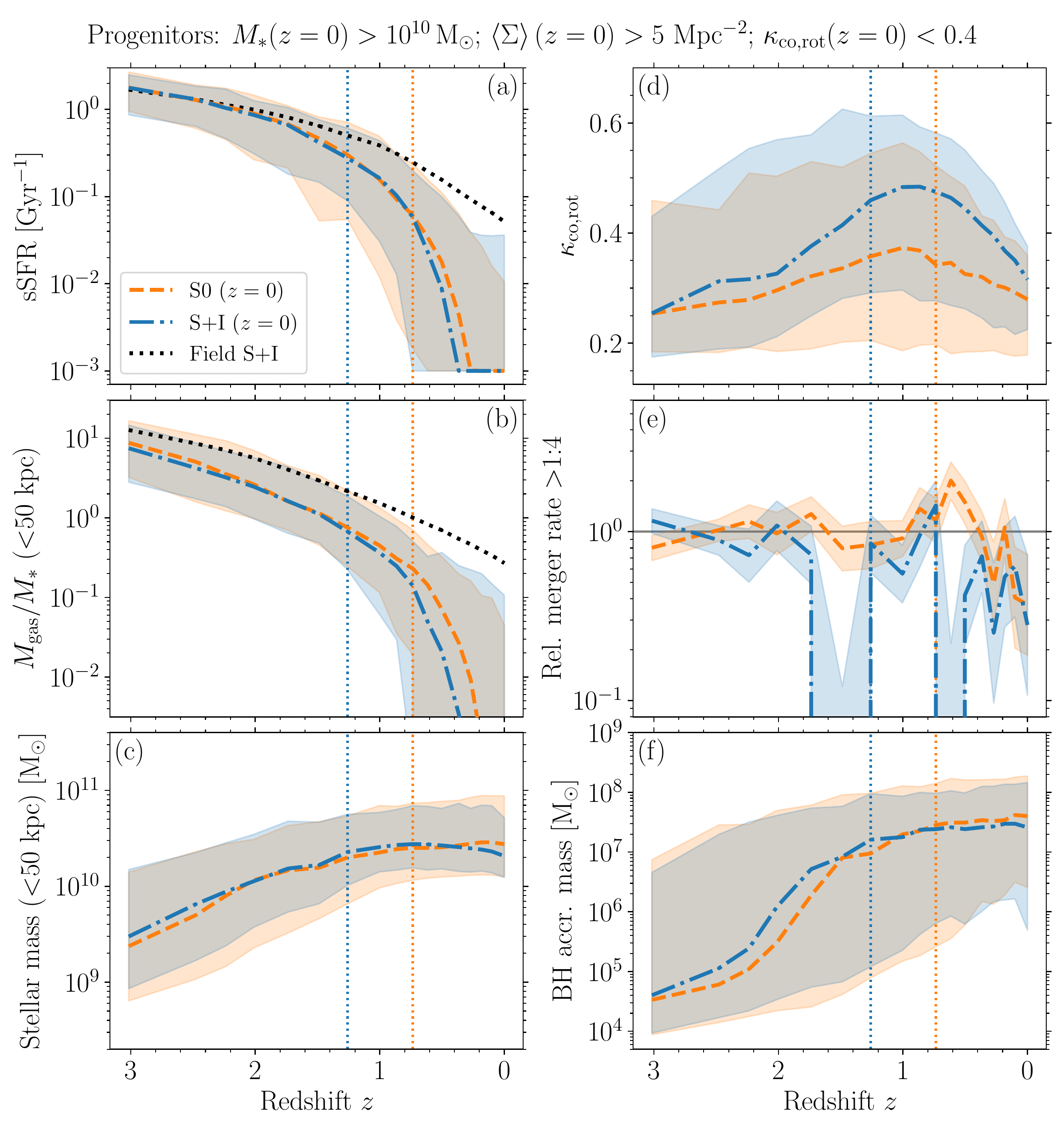}
  \caption{Redshift evolution of properties of main progenitors for galaxies residing in \textit{high} density environments ($\Sigma_3 > 5 \Mpc^{-2}$) and with $\kappaco < 0.4$ (spheroid-dominated) at $z=0$. Panels and line styles are as in Fig.~\ref{fig:progs_lowDens}. For reference, the black dotted lines show the sSFR and gas-to-stellar mass ratio evolution of low-density S+I galaxies in Fig.~\ref{fig:progs_lowDens}. Satellite S0 and S+I galaxies have median infall redshifts of $0.74$ and $1.26$, respectively, indicated by vertical dotted lines in each panel.}
  \label{fig:progs_highDens_lowKappa}
\end{figure*}

In Fig.~\ref{fig:progs_highDens} we compare the evolution of galaxies in high-density environments at $z=0$. Contrasting the results with Fig.~\ref{fig:progs_lowDens} reveals very different histories compared to galaxies in low densities, with the exception of the progenitors of elliptical galaxies which evolve similarly in both environments (the main differences being slightly higher masses, earlier stellar mass and black hole growth and an earlier peak in $\kappaco$ for high-density ellipticals).

The progenitors of late-type and S0 galaxies in high-density environments show very similar evolutionary histories, with the main difference being that late-type galaxies show slightly higher median $\kappaco$ at all redshifts (panel d). This might be explained by the slightly higher meger rates for S0 galaxies (panel e) and the larger tail (16\ts{th}-84\ts{th} percentiles) to lower $\kappaco$.
However, unlike low-density S0s, the progenitors of high-density S0s show \textit{lower} than average major merger rates (panel e) at $z < 0.5$, indicating infall into a larger group/cluster has resulted in lower merger rates.
We demonstrate this further in Fig.~\ref{fig:enviro_merger_rates} by comparing the relative major merger rates at $z=0$ as a function of galaxy morphology and environment.
All galaxy types show a correlation between merger rate and galactic density such that major mergers become less likely in high density environments, with the largest change in merger rate occurring for S0 galaxies.
This is (at least qualitatively) in agreement with expectations from $N$-body simulations that mergers are less frequent in galaxy clusters \citep{Ghigna_et_al_98}.
A possible exception to the trend is elliptical galaxies at the highest densities in the sample, which could be a result of being the central galaxies in groups/clusters (for $\sim$50 per cent of such galaxies, as compared to $<10$ per cent for S0 and late-type galaxies at the same density).
However, as the binomial uncertainties show, it is not a statistically significant result given the lower number of elliptical galaxies (22) in the highest density range.

In Fig.~\ref{fig:progs_highDens}, both late-type and S0 progenitors in high-density environments show a peak $\kappaco \sim 0.5$ (panel d) that is similar to the present-day value of low-density late-type galaxies (Fig.~\ref{fig:progs_lowDens}).
The redshift of peak $\kappaco$ also coincides with the typical infall redshift of $\approx 0.74$ for both galaxy types.
Therefore, given their similar sSFR and $\kappaco$, prior to group/cluster infall the progenitors of high-density S0s may be expected to appear visually similar to late-type galaxies.
That the decline in median $\kappaco$ at $z < 0.7$ is very similar between both late-type and S0 galaxies, despite the lower merger rate of late-type galaxies, suggests that mergers are unlikely to be the origin of the decline.
Other possible processes\footnote{\citet{Croom_et_al_21} showed that stellar population fading may lower the \textit{observed} stellar spin of galaxies. This process is unlikely to strongly affect $\kappaco$, which depends on the intrinsic angular momentum and not luminosity.} include heating following gas stripping (where gas makes a non-negligible contribution to the potential; e.g. at $z \sim 1$ both galaxy types have typical gas-to-stellar mass ratios $\approx 1$ within $50 \kpc$, panel b), harassment \citep{Moore_et_al_96} or numerical heating \citep[due to interactions between star particles and more massive dark matter particles,][]{Ludlow_et_al_19}.

Both late-type and S0 galaxies become (in the median) non-star forming by $z=0$ (panel a).
Unlike low-density environments (where over 70 per cent are central galaxies), $\approx$90 per cent of late-type and S0 galaxies in high-density environments are satellite galaxies, with a typical infall redshift of $\approx 0.75$ (i.e. time when a galaxy becomes a satellite in a halo, rather than the central galaxy).
The infall redshift coincides with the time at which the median sSFRs of S0 and late-type galaxies (panel a) experience a significant downturn relative to low-density S+I galaxies (black dotted lines in panel).
As shown in panel (b), for late-type and S0 galaxies it is not simply that the gas in the galaxies becomes non-star forming (i.e. through AGN feedback, as for elliptical galaxies which still have gas-to-stellar mass ratios of $\sim5$ per cent at $z=0$ but are largely non-star forming), but that most galaxies become completely gas free.
We suggest environmental processes such as tidal and ram pressure stripping \citep{Spitzer_and_Baade_51, Gunn_and_Gott_72} have removed the gas in most galaxies in high-density environments.
As shown by \citet{Pfeffer_et_al_22b}, the gas-stripping process in EAGLE galaxies occurs in an outside-in manner (leading to only galaxy centres remaining star forming; consistent with observations of cluster galaxies, e.g. \citealt{Koopmann_and_Kenney_04a, Koopmann_and_Kenney_04b, Reynolds_et_al_22}), with stronger stripping occurring in higher-mass group/cluster haloes.
However, AGN feedback could still play some role in these galaxies, even if stripping is responsible for complete quenching, as we note that high-density late-type and S0 galaxies have higher median accretion masses for black holes (panel f) than low-density late-type galaxies.
For example, one could conceive of a scenario in which the black holes experience a rapid growth phase prior to, or about the same time as, the galaxies become satellites and gas stripping takes place, possibly due to ram pressure driving gas inflows \citep{Bekki_09, Poggianti_et_al_17, Ricarte_et_al_20}.

To investigate this further we have divided the high-density S0 and S+I galaxies into high rotation (Fig.~\ref{fig:progs_highDens_highKappa}) and low rotation (Fig.~\ref{fig:progs_highDens_lowKappa}) galaxies at $\kappaco (z=0) = 0.4$ (i.e. intermediate between S0 and S+I galaxies in Fig.~\ref{fig:progs_lowDens}).
The cut in $\kappaco$ roughly divides the samples into two ($144/314$ S0 and $121/246$ S+I galaxies in the high rotation samples).
Following from the discussion in Section~\ref{sec:low-density}, we investigate whether low and high $\kappaco$ S0 and S+I galaxies in high-density environments also show different origins due to the presence or absence of AGN feedback.

In Fig.~\ref{fig:progs_highDens_highKappa} we show the progenitor evolution for high-density S0 and S+I galaxies with high rotation, $\kappaco (z=0.4) > 0.4$.
These galaxies generally do not experience a rapid black hole growth phase (panel f), and thus AGN feedback plays a limited role in their evolution.
Instead, both their sSFR (panel a) and gas-to-stellar mass ratios (panel b), along with $\kappaco$ (panel d), only begin to significantly decrease (relative to low density S+I galaxies) upon group/cluster infall (vertical dotted lines in each panel), indicating environmental gas stripping \citep{Spitzer_and_Baade_51, Gunn_and_Gott_72} as the origin of their decline.
At $z=0$, most S0 galaxies are almost completely quenched of star formation, while S+I galaxies have sSFRs $\approx 2.5$ times lower than field galaxies.
Relative to S+I galaxies, high $\kappaco$ S0 galaxies have slightly earlier infall times and lower sSFRs by $z=0$, but otherwise similar evolution in $\kappaco$, black hole accretion and merger rates (panels d, e, f).

In Fig.~\ref{fig:progs_highDens_lowKappa} we show the progenitor evolution for high-density S0 and S+I galaxies with low rotation, $\kappaco (z=0.4) < 0.4$, which show very different evolution compared to their high rotation counterparts.
Like the low-density S0 galaxies (Fig.~\ref{fig:progs_lowDens}), low-rotation S0 and S+I galaxies in high-density environments have typically undergone a rapid black hole growth phase, though at earlier times from $z \approx 2$-$1.5$ (panel f).
Their sSFRs (panel a) begin decreasing at similar redshifts ($z \approx 1.6$) as the rapid black hole growth, while the gas-to-stellar mass ratios (panel b) do not significantly decrease until after group/cluster infall.
Thus, AGN feedback initially begins to quench the galaxies, which is then completed by gas stripping after infall.
Interestingly, the evolution of $\kappaco$ is markedly different between the S0 and S+I galaxies (panel d), with S+I galaxy progenitors typically achieving much higher peak $\kappaco$ ($\approx 0.5$) compared to S0 galaxies ($\approx 0.35$).
Panel (c) shows the S+I galaxies experience some mass loss from $z \approx 0.8$ to $z=0$, which might indicate that tidal stripping/harassment effects on the discs is responsible for lowering $\kappaco$ \citep[c.f.][]{Moore_et_al_96, Moore_et_al_99, Bekki_and_Couch_11}.

\subsubsection{Elliptical galaxies}
\label{sec:ellipticals}

\begin{figure}
  \includegraphics[width=84mm]{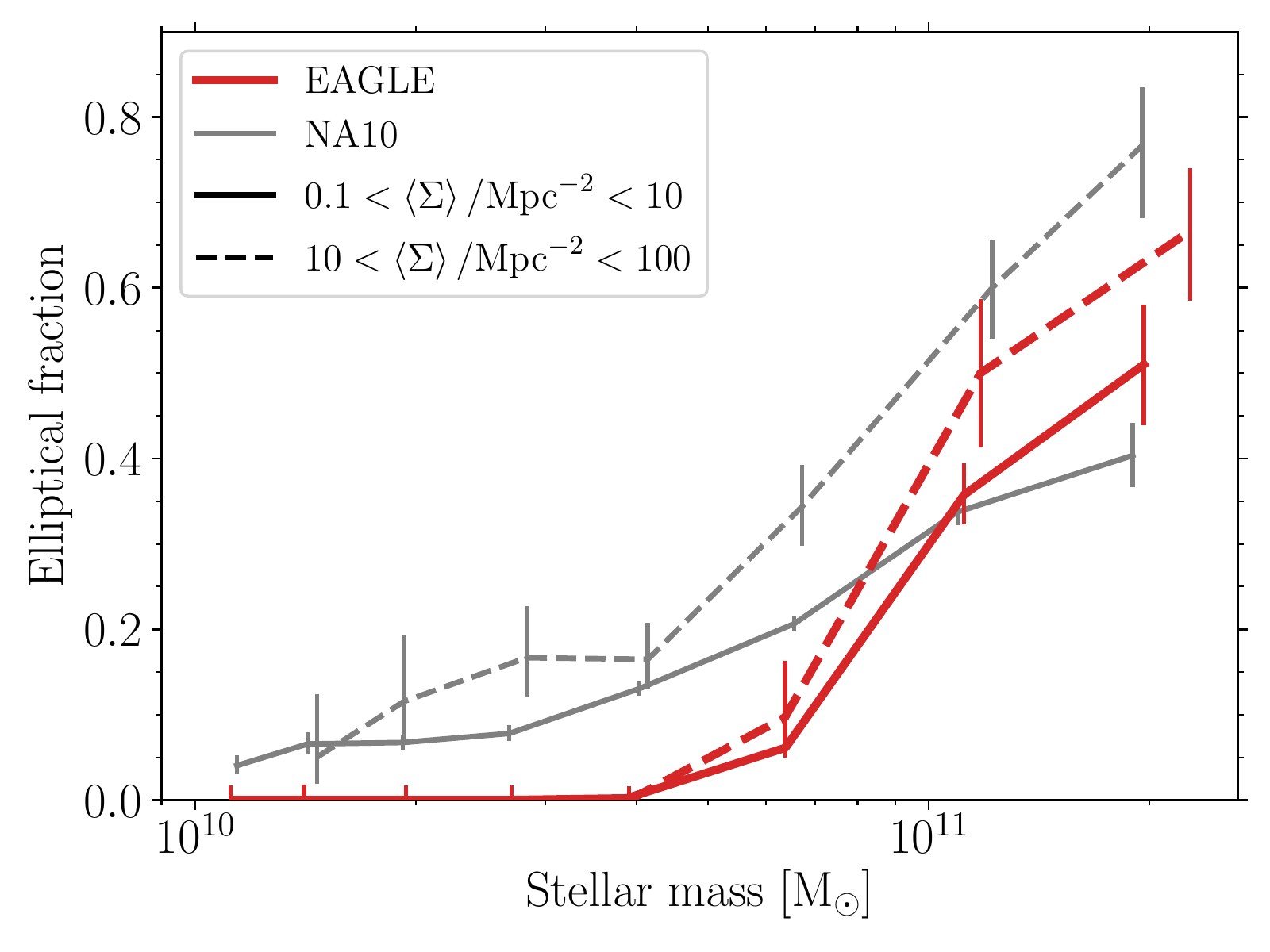}
  \includegraphics[width=84mm]{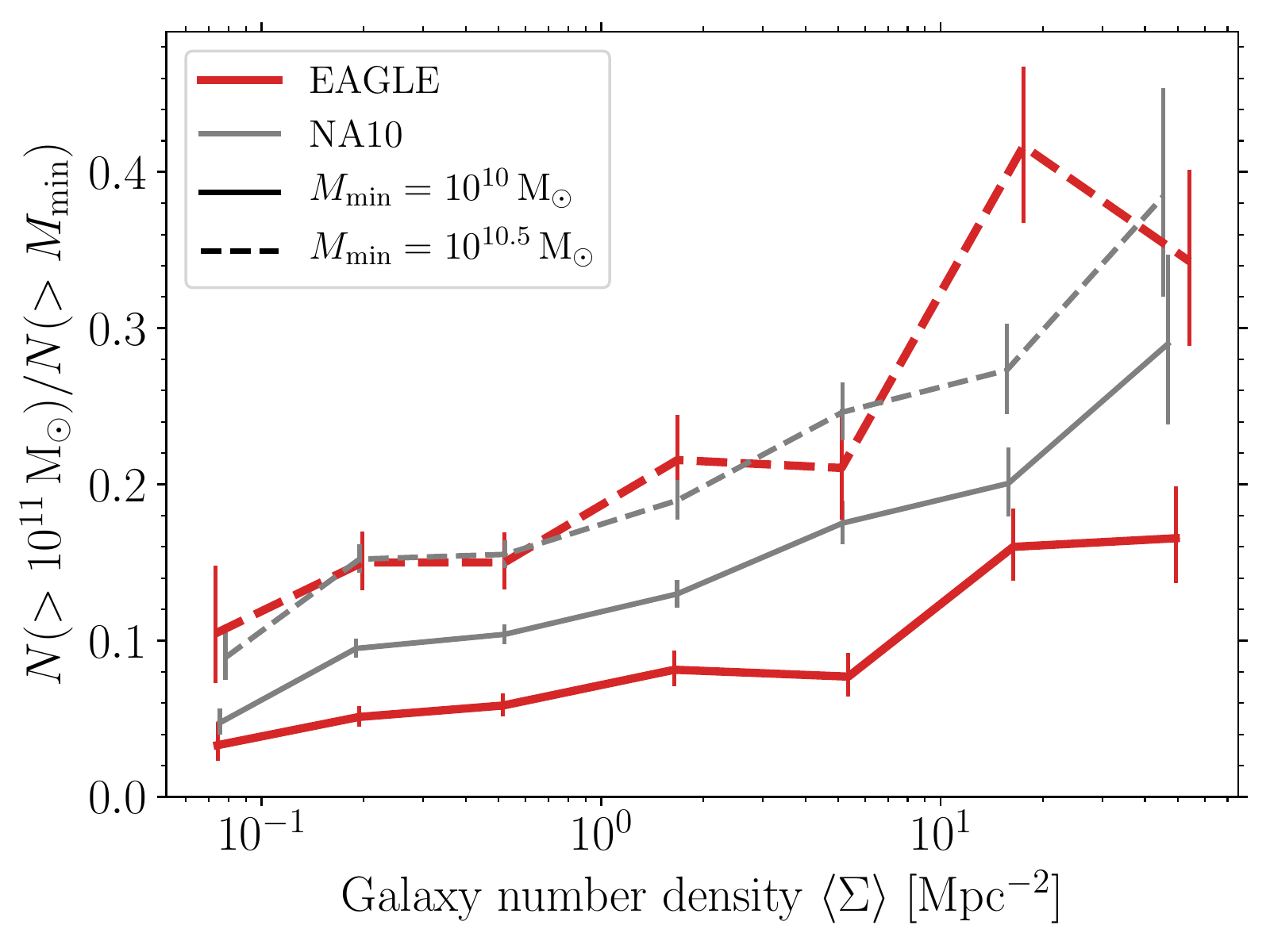}
  \caption{Comparison of elliptical and high-mass ($M_\ast > 10^{11} \Msun$) galaxy fractions between EAGLE (thick red lines) and \citet{Nair_and_Abraham_10} catalogue (thin grey lines). \textit{Top}: Elliptical fraction as a function of mass in low- (solid lines) and high-density environments (dashed lines), divided at $\SigAv = 10 \Mpc^{-2}$, where elliptical fraction is highest (Fig.~\ref{fig:morph-dens}). \textit{Bottom}: Fraction of galaxies more massive than $M_\ast > 10^{11} \Msun$ as a function of environment. Solid lines show fraction for a minimum stellar mass limit of $10^{10} \Msun$ and dashed lines show a minimum limit of $10^{10.5} \Msun$. In both panels we show bins in mass or number density with more than 20 galaxies.}
  \label{fig:elliptical_fracs}
\end{figure}

In Sections~\ref{sec:low-density} and \ref{sec:high-density} we only briefly discussed the formation of elliptical galaxies relative to S0 and late-type galaxies.
As shown in Figures~\ref{fig:progs_lowDens} and \ref{fig:progs_highDens}, elliptical galaxies experience early star-formation quenching (through AGN feedback) due to rapid black hole growth at early times ($z \gtrsim 2$) and have the lowest rotational support due to their high galaxy merger rates (from $z \lesssim 1.5$).
In low-density environments the majority of ellipticals ($\approx 90$ per cent) are central galaxies in their dark matter halo, compared with only $\approx 44$ per cent in high-density environments.
However, ellipticals which are classed as satellite galaxies in high-density environments have typical infall redshifts of $\approx 0.27$, much later than S0 and late-type galaxies, but consistent with the drop in merger rate at $z < 0.3$ in panel (e) of Fig.~\ref{fig:progs_highDens} and the later assembly times of more massive dark matter haloes \citep[e.g.][]{De_Lucia_et_al_06, Qu_et_al_17}.

In Fig.~\ref{fig:elliptical_fracs} we investigate the origin of the increasing fraction of elliptical galaxies with density.
As the top panel of Fig.~\ref{fig:elliptical_fracs} shows, for lower-mass galaxies ($M_\ast \sim 10^{10} \Msun$), which dominate galaxy numbers overall due to the steep galaxy mass function, there is almost no difference in elliptical fraction for low ($\SigAv < 10 \Mpc^{-2}$) and high ($\SigAv > 10 \Mpc^{-2}$) densities in the \citet{Nair_and_Abraham_10} catalogue.
Given the issues identifying low-mass elliptical galaxies in the EAGLE simulation (Section~\ref{sec:props}), if low-mass galaxies drove the morphology-density relation for elliptical galaxies then we would expect a flat relation for EAGLE in Fig.~\ref{fig:morph-dens}, which is not what is found.

Instead, as the bottom panel of Fig.~\ref{fig:elliptical_fracs} shows, the fraction of high-mass galaxies ($M_\ast > 10^{11} \Msun$, compared with all galaxies greater than $10^{10}$ and $10^{10.5} \Msun$) increases with galaxy density, likely to be driven by mass segregation (through dynamical friction) and tidal disruption/merging of lower mass galaxies \citep[c.f.][]{Whitmore_Gilmore_and_Jones_93}.
This, combined with the elliptical fraction being a strong function of galaxy mass, and a secondary dependence on density for high mass galaxies (top panel, i.e. in dense environments more high-mass galaxies are elliptical), drives the morphology-density relation for elliptical galaxies.

\citet{Whitmore_Gilmore_and_Jones_93} found that $\sim$55 per cent of galaxies at the centres of clusters are elliptical, regardless of local density, and showed that the morphology-radius relation appears to be the more fundamental correlation.
Analogously, \citealt{Houghton_15} found the elliptical fraction rises faster with density in lower density clusters \citep[confirming trends found when only considering slow-rotating galaxies][]{Cappellari_et_al_11, DEugenio_et_al_13, Houghton_et_al_13, van_de_Sande_et_al_21}.
\citet{van_der_Wel_et_al_10} argued both the elliptical morphology-density and morphology-radius correlations are due to cluster centres hosting more massive galaxies, which are more often ellipticals, and indeed we find a similar result for the morphology-density relation with the EAGLE simulation.

\section{Discussion and summary}
\label{sec:summary}

\subsection{The galaxy morphology-density relation and the origin of S0 galaxies}

Much of the discussion on the origin of the morphology-density relation is closely tied with discussion on the origin of S0 galaxies.
Recent works have shown that S0 galaxies do not appear to be a homogeneous population, but that their formation may vary with environment.
Indeed the form of the morphology-density relation itself, with a non-zero fraction of S0 galaxies at low densities, suggests environmentally-dependent mechanisms (like stripping) cannot be the only origin \citep{Dressler_80a, Postman_and_Geller_84, van_der_Wel_et_al_10}.
In particular, S0 galaxies in clusters appear more rotationally supported than those in field \citep{Coccato_et_al_20, Deeley_et_al_20} and, similarly, quiescent galaxies are more disc-dominated in denser environments (\citealt{van_der_Wel_et_al_10}; which is also found for the EAGLE simulation, \citealt{Correa_et_al_17}).
This appears to be consistent with a transition from (largely) a merger origin for field S0s to a stripped spiral origin for cluster S0s \citep[this work]{Deeley_et_al_21}.
Low-mass ($M_\ast < 10^{10} \Msun$) field S0 galaxies may have a different origin again to higher mass galaxies \citep{Fraser-McKelvie_et_al_18, Dolfi_et_al_21}, which might be expected if they are too low mass for their formation to be affected by black hole feedback \citep[e.g.][]{Crain_et_al_15}.

At face value, that S0 galaxies in clusters are more rotationally supported than field galaxies appears at odds with the lower disc-to-bulge ratios of S0 galaxies compared to spiral galaxies \citep{Dressler_80a}.
Comparing to the disc-to-bulge ratios of S0 galaxies from \citet{Dressler_80a} over a similar range in $\Sigma$ ($1$-$50 \Mpc^{-2}$) indeed shows that $\kappaco$ \citep[which correlates with disc-to-bulge ratio,][]{Thob_et_al_19} is relatively constant and lower than that of disc galaxies, while the drop to lower $\kappaco$, merger-dominated formation occurs at $\lesssim 1 \Mpc^{-2}$ (Fig.~\ref{fig:props-dens}).

A merger origin (both major and minor) for S0 galaxies in low-density environments has of course been discussed in many works \citep[e.g.][]{Bekki_98, Eliche-Moral_et_al_12, Eliche-Moral_et_al_13, Borlaff_et_al_14, Querejeta_et_al_15, Tapia_et_al_17, Diaz_et_al_18, Deeley_et_al_21}.
One of the key findings of this work (Section~\ref{sec:low-density}) is that, though mergers are responsible for the transformation to more dispersion-dominated systems, it is the black hole feedback induced by mergers that results in their lower star-formation rates compared to late-type galaxies \citep[e.g.][]{Bait_et_al_17}.

\subsection{Summary}

In this paper, we present the first analysis of the galaxy morphology-density relation in a cosmological hydrodynamical simulation.
We used a CNN trained on observed galaxies \citep{Cavanagh_et_al_22} to perform `visual' morphological classification of galaxies in the EAGLE simulation \citep{Schaye_et_al_15, Crain_et_al_15}.
Our main findings from the work are as follows:

We first demonstrated in Section~\ref{sec:props} that EAGLE reproduces the observed morphology-galaxy mass relations (Fig.~\ref{fig:morph-mass}):
Late-type and elliptical fractions are a strong function of galaxy mass, with late-type fraction decreasing and elliptical fraction increasing with mass, while the fraction of lenticular galaxies is relatively constant ($\sim30$-$40$ per cent).
However almost no galaxies with stellar masses $<5 \times 10^{10} \Msun$ are classified as elliptical, compared to observed fractions of $\sim 5$-$10$ per cent.
This issue likely stems from the insufficiently concentrated galaxy profiles due to numerical effects \citep{de_Graaff_et_al_21}, resulting in slightly elevated S0 fractions.

We also showed in Fig.~\ref{fig:type-props} that the physical properties for simulated galaxies of each morphological type are in line with those expected from observations \citep[e.g.][]{Cortese_et_al_16, Bait_et_al_17, Falcon-Barroso_et_al_19}.
Late-type galaxies are typically star forming (high sSFR) and disc dominated (high $\kappaco$), elliptical galaxies are spheroidal (low $\kappaco$) and non-star forming (low sSFR), while S0 galaxies are intermediate in both properties.

In Section~\ref{sec:morph-dens} we showed, for the first time, that simulations can reproduce the observed galaxy morphology-density relation (Fig.~\ref{fig:morph-dens}).
The EAGLE model is therefore ideal to test the origin of the relation.
However the simulation does not reach the highest densities probed by observations (i.e. centres of rich galaxy clusters) due to the limited simulation volume.

In Sections~\ref{sec:props-enviro} and \ref{sec:enviro_processes} we investigated the physical processes that shape the morphology-density relation in the simulations.
We found three key drivers of the relation:
\begin{itemize}
  \item Transformation of disc-dominated galaxies from late-type to S0 through gas stripping in higher density environments \citep[c.f.][]{Gunn_and_Gott_72, Larson_et_al_80, Bekki_et_al_02}, which accounts for the decreasing late-type fraction at higher densities.
  \item An origin for S0 galaxies that transitions from merger-dominated to stripping-/starvation-dominated from low to high-density environments \citep[c.f.][]{Deeley_et_al_21}, which accounts for the increasing S0 fraction with density. In addition, we showed that AGN feedback driven by mergers may play a significant role in the formation of field S0s \citep[c.f.][]{Davies_Pontzen_and_Crain_22}.
  \item The increasing fraction of high-mass galaxies, that are more likely ellipticals, which drives the rapid increase in elliptical galaxies at higher densities \citep[c.f.][]{van_der_Wel_et_al_10}.
\end{itemize}

\subsection{Future directions}

In this work we have not investigated the redshift dependence of the morphology-density relation.
Most high redshift observations focus on rich galaxy clusters \citep[][though see \citealt{van_der_Wel_et_al_07, Shimakawa_et_al_21}]{Dressler_et_al_97, Couch_et_al_98, Fasano_et_al_00, Treu_et_al_03, Postman_et_al_05, Smith_et_al_05}, while the EAGLE volume ($100^3 \cMpc^3$) is too small to contain very massive clusters.
Therefore, in future work we will extend our analysis to the Cluster-EAGLE/Hydrangea simulations \citep{Bahe_et_al_17, Barnes_et_al_17}, a set of 30 zoom-in simulations of galaxy clusters ($M_{200} \sim 10^{14}$-$10^{15.4} \Msun$).
This will enable us to compare the redshift evolution of the morphology-density relation, the cluster morphology-radius relation \citep{Dressler_80a, Whitmore_Gilmore_and_Jones_93} and the Butcher-Oemler effect \citep[][]{Butcher_and_Oemler_78, Butcher_and_Oemler_84, Couch_et_al_94} with observations in similar environments.

We also have not discussed the neutral hydrogen (\HI) morphology-density relation \citep{Serra_et_al_12} and its connection to the optical morphology-density relation.
In principle, the combination of optical and \HI\ data may aid in determining the origin of individual early-type galaxies, rather than examining population-averaged results as we have done in this work.
In future, the Widefield ASKAP L-band Legacy All-sky Blind surveY (WALLABY) survey with the Australian Square Kilometre Array Pathfinder (ASKAP) telescope \citep{Koribalski_et_al_20} is expected to provide \HI\ properties for more than $10^5$ galaxies, significantly expanding the capability to study the morphology-density relation in the nearby Universe.

\section*{Acknowledgements}

We thank the referee for a very constructive report which improved the paper.
This research was supported by the Australian government through the Australian Research Council's Discovery Projects funding scheme (DP200102574).
This work used the DiRAC Data Centric system at Durham University, operated by the Institute for Computational Cosmology on behalf of the STFC DiRAC HPC Facility (\url{www.dirac.ac.uk}). This equipment was funded by BIS National E-infrastructure capital grant ST/K00042X/1, STFC capital grants ST/H008519/1 and ST/K00087X/1, STFC DiRAC Operations grant ST/K003267/1 and Durham University. DiRAC is part of the National E-Infrastructure.

\section*{Data Availability}

The data underlying this article will be shared on reasonable request to the corresponding author.
All data from the EAGLE simulations (including galaxy catalogues, merger trees and particle data) is publicly available \citep{McAlpine_et_al_16} at \href{http://www.eaglesim.org/database.php}{http://www.eaglesim.org/database.php}.



\bibliographystyle{mnras}
\bibliography{bibliography}



\appendix

\section{CNN accuracy testing}
\label{app:CNN_test}

To test the accuracy of the CNN (Section~\ref{sec:CNN}) when applied to simulated galaxy images, in this appendix we compare the CNN classifications with human classifications of the same galaxies.
As previously, the CNN classifications were performed using a single face-on image for each galaxy.
However, both face-on and edge-on images of the galaxies were used for the human visual classifications to help determine the `true' morphology of each galaxy.
In particular, we aim to distinguish between spheroid-dominated (E) and disc-dominated (S0) early-type galaxies to compare with the classifications from the CNN.
The issue of distinguishing E and S0 galaxies from single images also exists in classifications of observed galaxies, where galaxy kinematics are necessary to distinguish between rotating and non-rotating early-type galaxies \citep[e.g. see][]{Cappellari_et_al_07, Emsellem_et_al_07, Emsellem_et_al_11}.

Using the process described in Section~\ref{sec:analysis}, images were generated for galaxies with $M_\ast > 10^{10} \Msun$ in the $z=0$ snapshot of the EAGLE RefL050N752 simulation of a $50^3 \cMpc^3$ volume \citep[which uses an identical resolution and subgrid model parameters as the largest EAGLE volume,][]{Schaye_et_al_15}.
This gives us a volume-limited sample of 475 galaxies, which are independent from the simulation used for the main results of the paper (EAGLE RefL100N1504).

Four of the authors (JP, WJC, MJD, DAF) classified each galaxy by visual inspection under an agreed set of criteria.
The classification tree can be summarised as:
\begin{itemize}
  \item Does the galaxy have prominent spiral or star formation features (i.e. is it a late-type galaxy)?
  \begin{itemize}
    \item If yes, does the galaxy have a regular disc component?
    \begin{itemize}
      \item If yes $\Rightarrow$ \textbf{Spiral (Sp)}.
      \item If no $\Rightarrow$ \textbf{Irregular (Irr)}.
    \end{itemize}
    \item If no, does the galaxy have a regular disc component?
    \begin{itemize}
      \item If yes $\Rightarrow$ \textbf{Lenticular (S0)}.
      \item If no, does the galaxy have an irregular morphology?
      \begin{itemize}
        \item If yes $\Rightarrow$ \textbf{Irregular (Irr)}.
        \item If no $\Rightarrow$ \textbf{Elliptical (E)}.
      \end{itemize}
    \end{itemize}
  \end{itemize}
\end{itemize}
A representative set of images (around 20 of each type) were selected and classified by all four authors to serve as a reference.
The full set of galaxies were then classified individually by each of the four authors.
This classification scheme uses fewer categories than that of \citet{Nair_and_Abraham_10}, who used the Hubble T-Type system. However, it is intended to be broadly consistent with the down-sampled categories used in the CNN analysis (e.g. spirals are defined as all types from Sa to Sm).

\begin{figure*}
  \centering
  \includegraphics[width=0.495\textwidth]{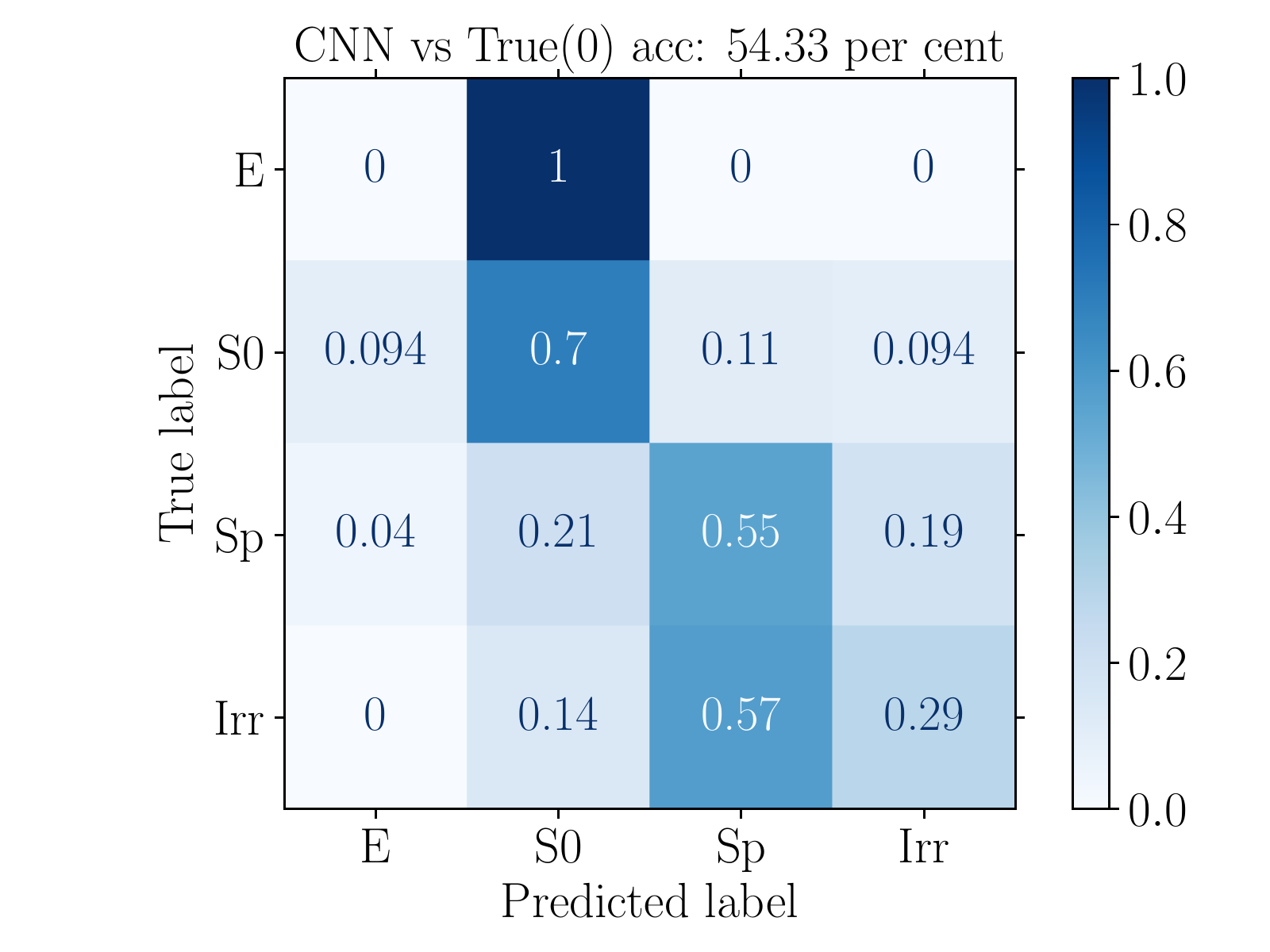}
  \includegraphics[width=0.495\textwidth]{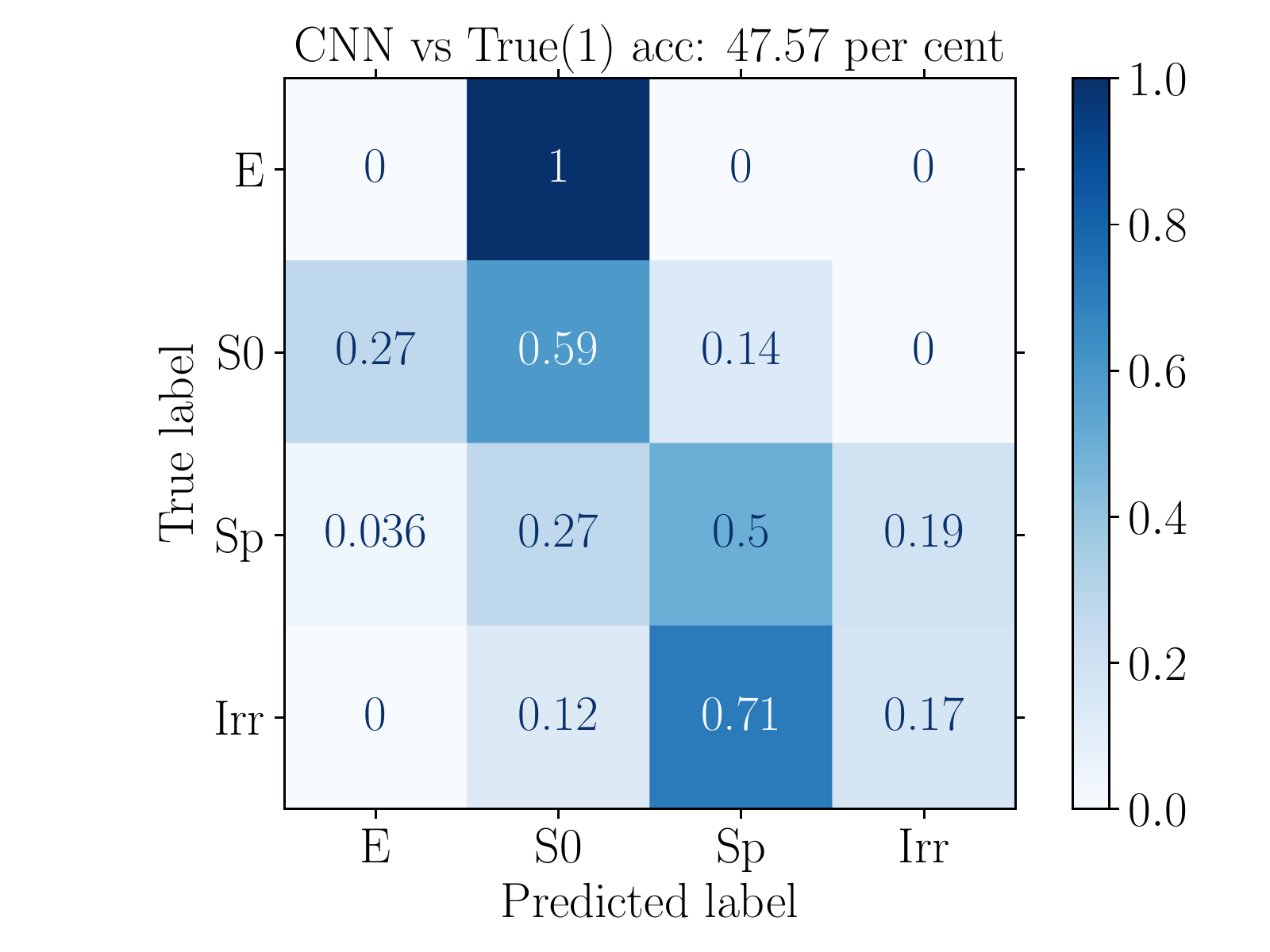}
  \includegraphics[width=0.495\textwidth]{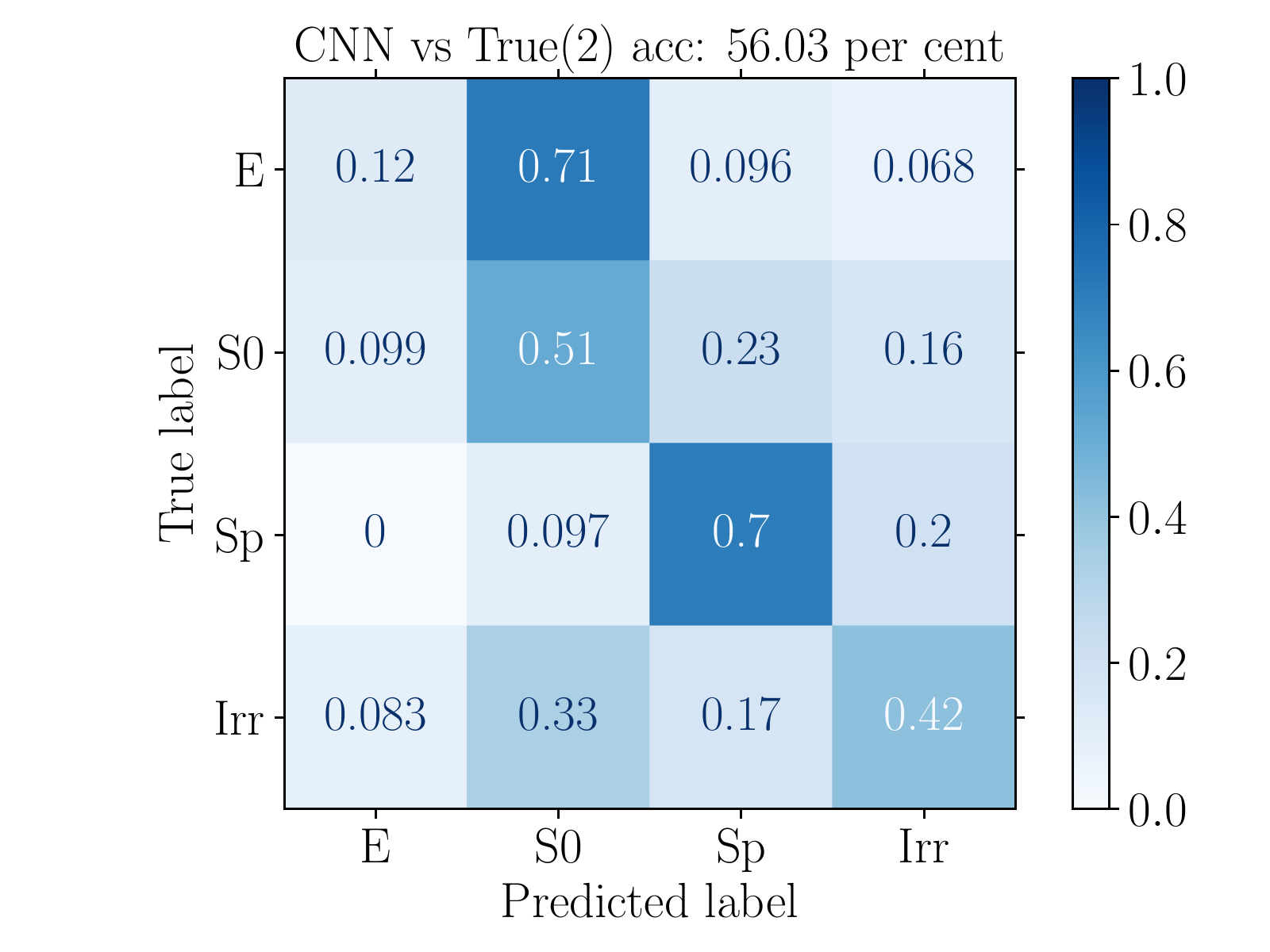}
  \includegraphics[width=0.495\textwidth]{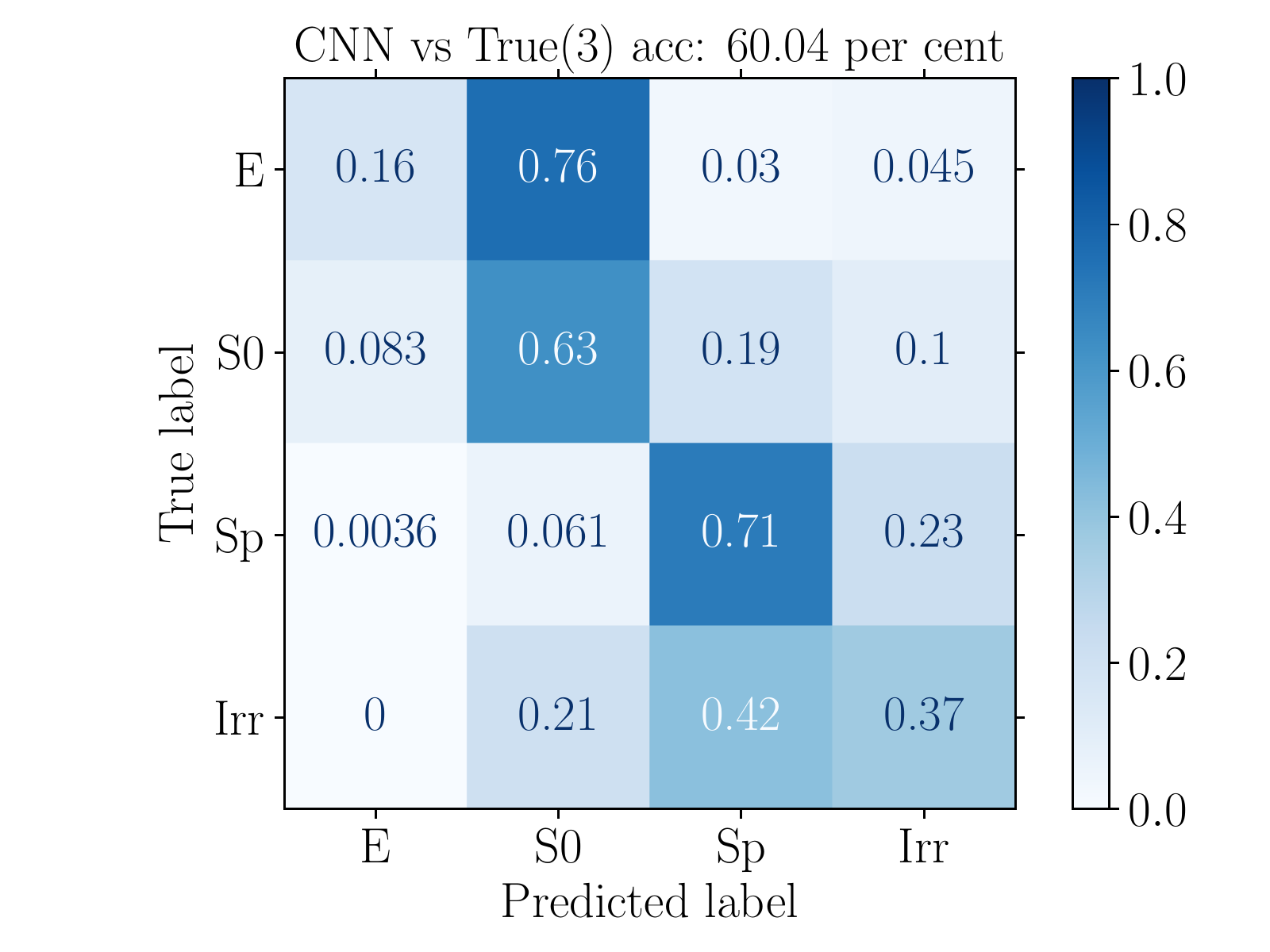}
  \caption{Confusion matrices for the CNN classifications (predicted labels) compared with each set of visual (`true') classifications of the galaxies from EAGLE RefL050N752. The overall accuracies are indicated in the title of each sub-panel.}
  \label{fig:confusions}
\end{figure*}

Fig.~\ref{fig:confusions} shows the confusion matrices for the CNN compared against the four sets of visual classifications.
Generally, galaxies classified as early- (E, S0) or late-type (Sp, Irr) in visual classifications are also classified as such by the CNN.
However, visually classified ellipticals tend to be classified as lenticular galaxies by the CNN, while visually classified irregulars tend to be classified as spiral galaxies by the CNN. 
The classification accuracy for spiral and lenticular galaxies is in the range of 50-70 per cent, depending on set of visual classifications.

\begin{figure*}
  \centering
  \includegraphics[width=\textwidth]{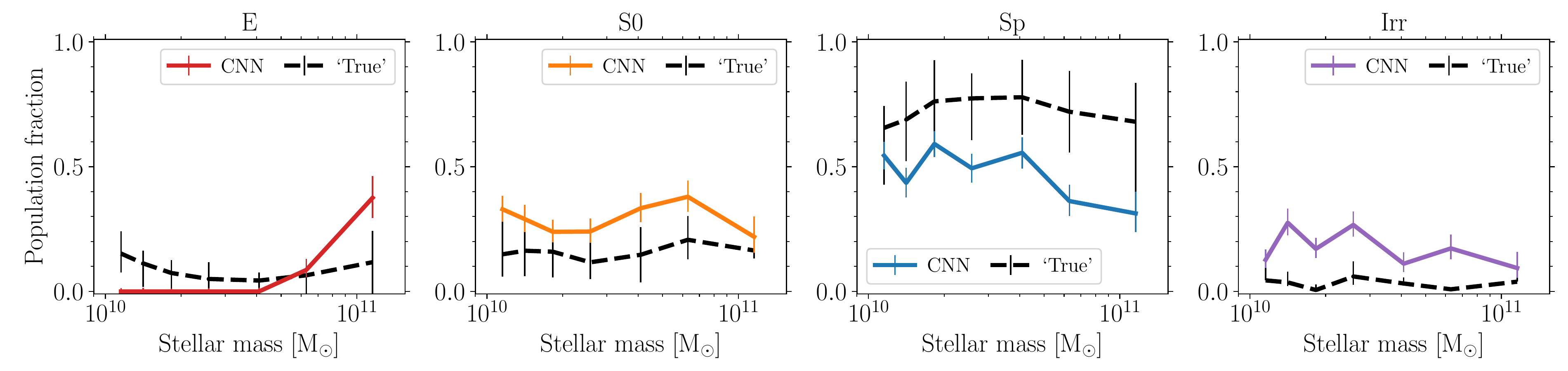}
  \caption{Dependence of morphology on galaxy stellar mass for galaxies in the EAGLE RefL050N752 simulation. Panels from left to right show the fraction of elliptical (E), lenticular (S0), spiral (Sp) and irregular (Irr) galaxies as a function of galaxy mass. Coloured lines show classifications from the CNN, with errorbars showing the uncertainties from binomial statistics. Black dashed lines show the average morphological fractions of four visual classifications of the same galaxies, with errorbars showing the full range of individual classification. Each mass bin contains more than 30 galaxies.}
  \label{fig:L50_comparison}
\end{figure*}

\begin{figure}
  \centering
  \includegraphics[width=84mm]{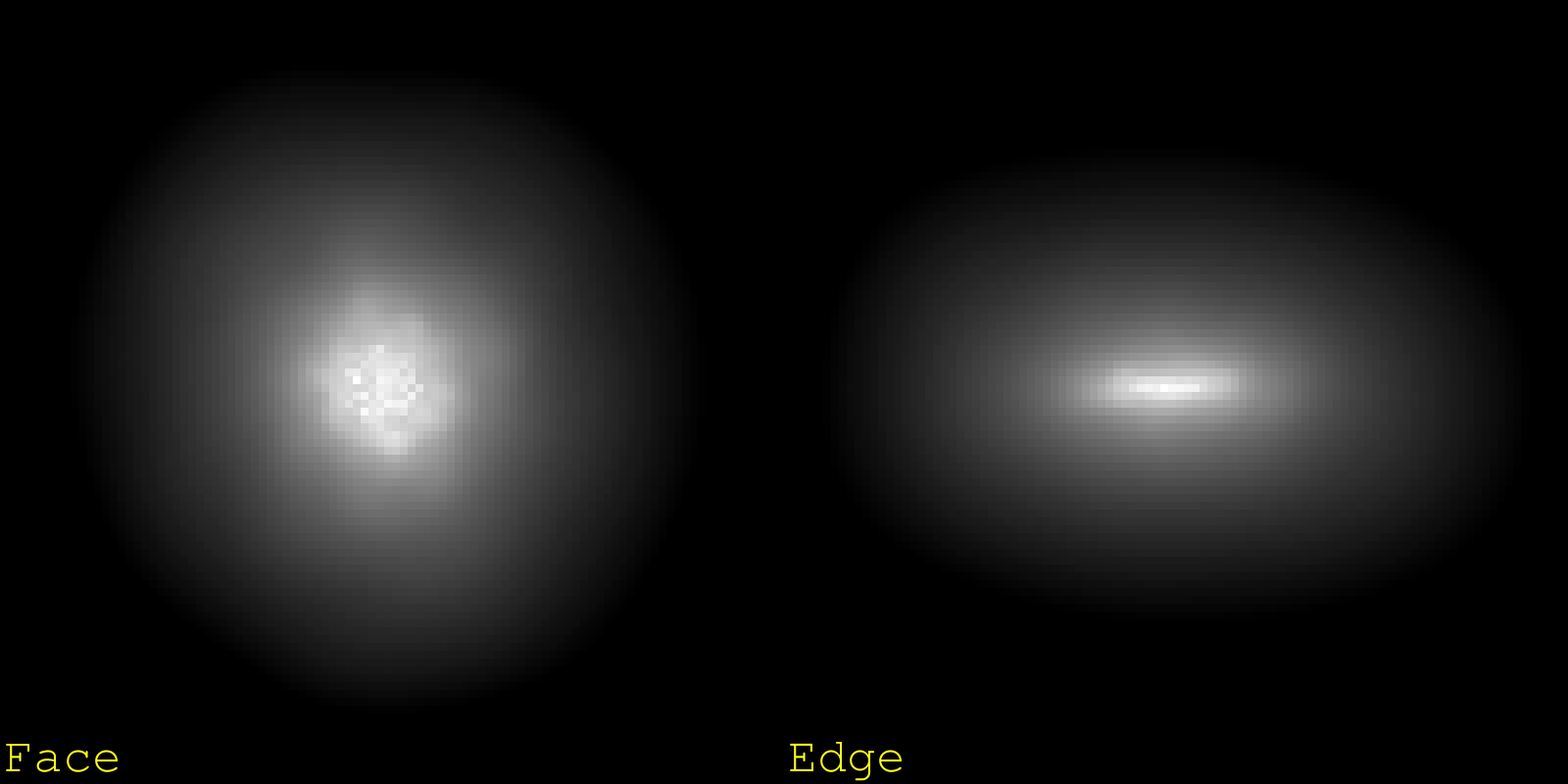}
  \caption{Example of a galaxy where visual classifications were evenly split between S0 and Sp. This galaxy was classified by the CNN as an S0. The left and right panels show the face-on and edge-on projections, respectively.}
  \label{fig:2353168}
\end{figure}

We investigate the classification accuracy further in Fig.~\ref{fig:L50_comparison} by comparing the galaxy morphology-mass relations (as in Fig.~\ref{fig:morph-mass}).
The coloured lines show the results for the CNN (with errorbars showing binomial uncertainties), while black dashed lines and errorbars show the mean and range for visual classifications.
Between individual authors there exists some systematic offset in the classified fractions of early- and late-type galaxies, which is most evident in the range of spiral fractions (as well as Fig.~\ref{fig:confusions}).
This reflects the difficulty in classifying galaxies that may be intermediate between S0 and Sp (e.g. Fig.~\ref{fig:2353168}).

\begin{figure}
  \centering
  \includegraphics[width=84mm]{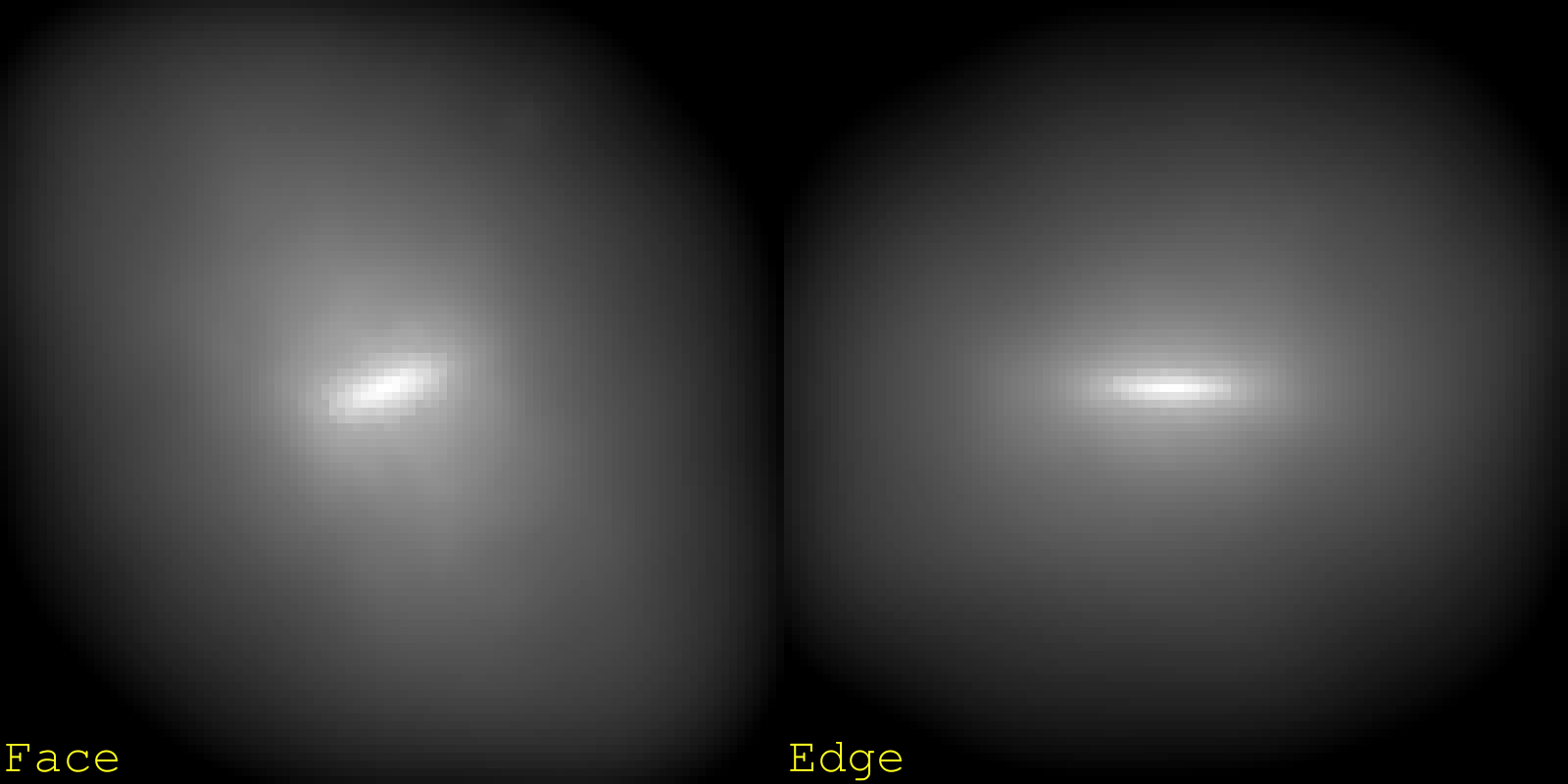}
  \caption{Example of a galaxy classified as an elliptical by the CNN and as a S0 by visual classification. The left and right panels show the face-on and edge-on projections, respectively.}
  \label{fig:1048971}
\end{figure}

\begin{figure}
  \centering
  \includegraphics[width=84mm]{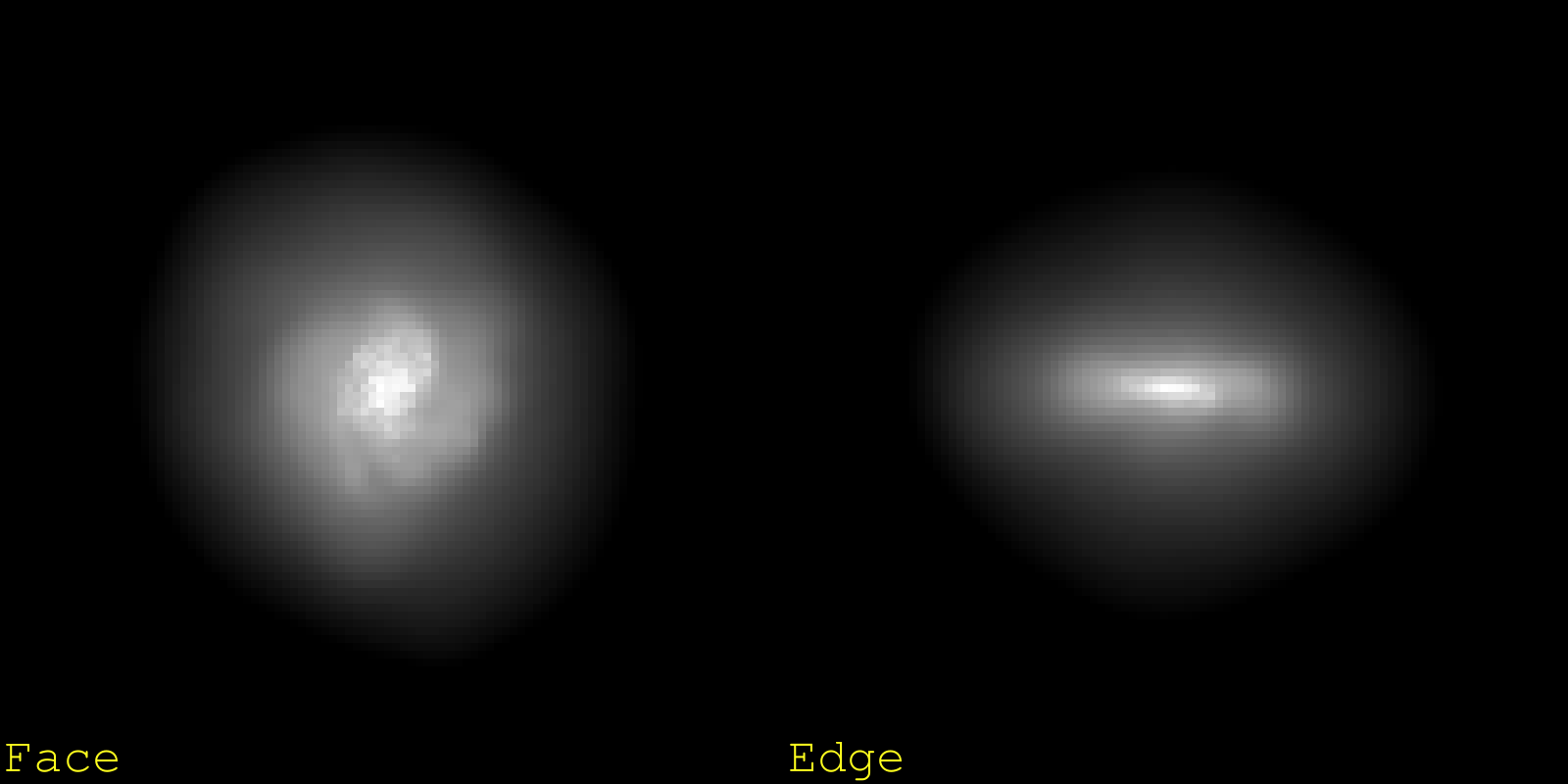}
  \caption{Example of a galaxy classified as an irregular by the CNN and as a spiral by visual classification. The left and right panels show the face-on and edge-on projections, respectively.}
  \label{fig:434818}
\end{figure}

\begin{figure}
  \centering
  \includegraphics[width=84mm]{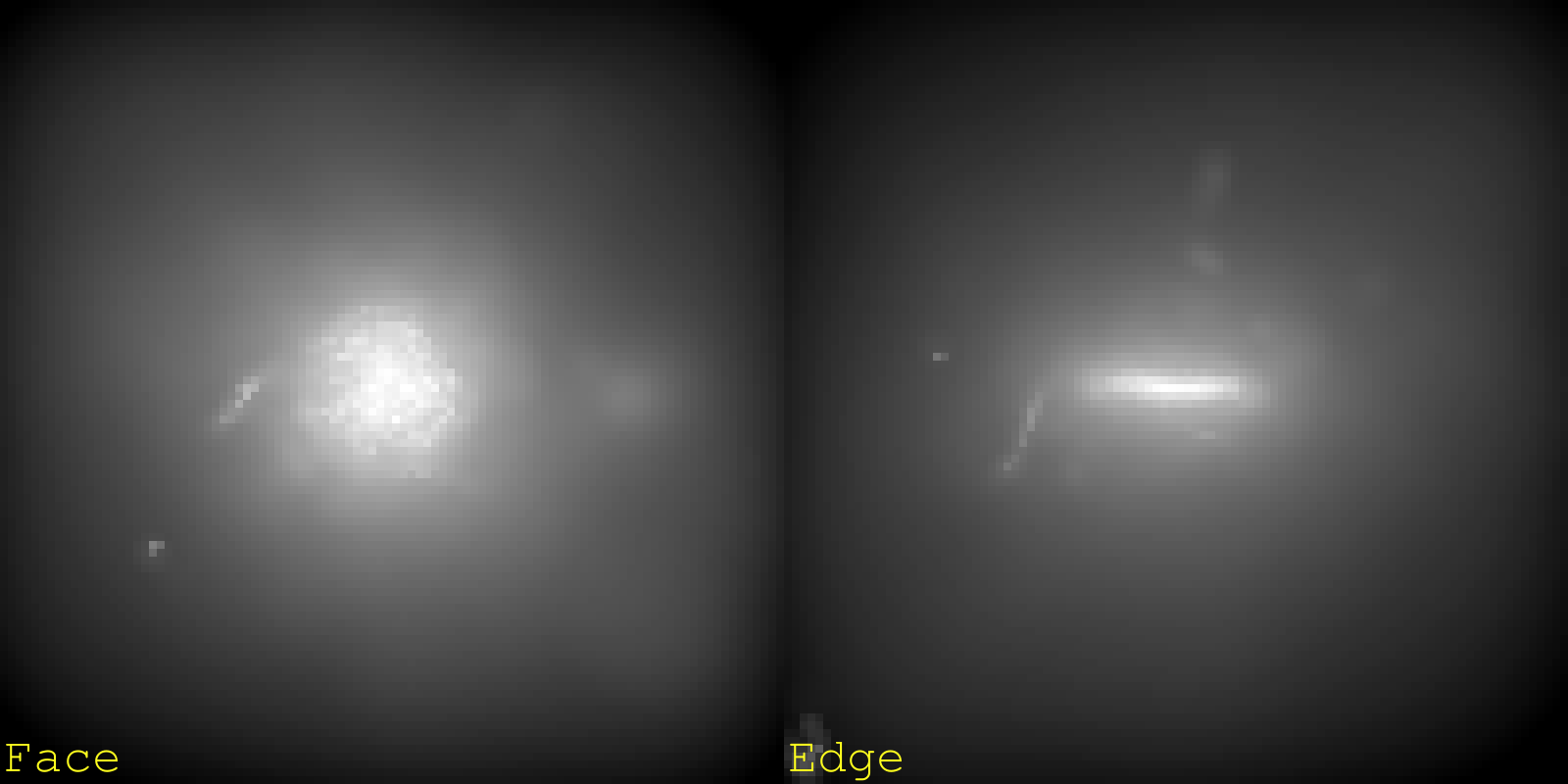}
  \caption{Example of a high mass ($M_\ast \approx 10^{11} \Msun$) galaxy classified as a S0 by the CNN and as a late-type by visual classification. The left and right panels show the face-on and edge-on projections, respectively.}
  \label{fig:2520223}
\end{figure}

Below we discuss the key deviations between CNN and visual classifications:
\begin{itemize}

\item \textit{No lower-mass ($M_\ast < 5 \times 10^{10} \Msun$) galaxies are classified by the CNN as elliptical}.
Around 9 per cent of such galaxies are classified as elliptical by visual inspection.
These galaxies are generally classified by the CNN as S0 galaxies ($>80$ per cent).
In Section~\ref{sec:props} we suggest that this issue may be due to insufficient concentration of spheroidal galaxies due to numerical effects.
To test this, we fit \citet{Sersic_63} profiles to the $g$-band profiles of each galaxy (using radially-averaged face-on projections) to compare the \Sersic\ indices ($n_s$).
For galaxies with $M_\ast < 5 \times 10^{10} \Msun$, we find those visually classified (in the majority) as E and S0 have median $n_s \approx 2.2$ and $1.5$, respectively.
In contrast, high mass galaxies ($M_\ast > 5 \times 10^{10} \Msun$) classified by the CNN as E and S0 have median $n_s \approx 3.8$ and $2.7$, respectively.
Therefore, lower-mass Es have less concentrated profiles than high mass S0s, which may be leading the CNN to classify the low-mass Es as S0s.

\item \textit{The CNN classifies more high-mass galaxies as elliptical than visual classifications.}
This difference is a result of the visual classifications making use of edge-on projections to distinguish genuine elliptical galaxies from lenticular galaxies.
We show an example of such a galaxy classified as an elliptical by the CNN and a lenticular in visual classifications in Fig.~\ref{fig:1048971}.
This issue is related (though not identical) to the finding of a lower fraction of slow-rotating (i.e. genuinely spheroidal) galaxies in `kinematic' morphologies compared to the fraction of elliptical galaxies in visual morphologies \citep[e.g.][]{Emsellem_et_al_07, Emsellem_et_al_11, Cappellari_et_al_11}.
However, many of the visually classified elliptical galaxies may still be fast-rotating galaxies.
For the visual classifications, 25-40 per cent of early-type galaxies were classified as elliptical.
This is much higher than the $\approx14$ per cent of slow-rotating early-type galaxies found for observed galaxies \citep{Emsellem_et_al_11}, despite the fraction of slow-rotating galaxies in EAGLE reasonably agreeing with observed fractions \citep{Lagos_et_al_18b}. 

\item \textit{The CNN classifies more galaxies as irregular than visual classifications.} Around 82 per cent of these galaxies are classified visually as spirals and 5 per cent classified as `true' irregulars (averaging over the four visual classifications). Taken together, $\approx 92$ per cent of galaxies are correctly identified by the CNN as late-type (S+I). The difference in irregular classifications is therefore likely due to difference in definition.
The CNN tends to classify only galaxies with strong spiral features as Sp and galaxies with weaker features as Irr.
In contrast, visual classifications also labelled galaxies with weaker star-formation features as Sp (e.g. Fig.~\ref{fig:434818}), while the irregular classification is reserved for star-forming galaxies without prominent discs and galaxies with irregular shapes.

\item \textit{More high-mass galaxies are classified as late-type (S+I) in visual classifications than by the CNN.} For galaxies with $M_\ast > 10^{11} \Msun$ the average late-type fraction in visual classifications is $0.7$, compared to $0.4$ for the CNN. However the CNN is consistent within the full range of visual classifications, given the systematic differences between individual sets of classifications (as stated above). For high-mass galaxies in particular, classification is often complicated by compact star-forming discs, and whether more weight should be given to the central region or the lack of features in the majority of the galaxy (e.g. Fig.~\ref{fig:2520223}). Such galaxies tend to be labelled by the CNN as S0.

\item \textit{The CNN classifies more galaxies as lenticular than visual classifications.} 
This difference is a combination of the factors discussed above.
At low masses ($M_\ast \lesssim 5\times 10^{10} \Msun$) spheroidal galaxies are classified by the CNN as S0, accounting for 7-37 per cent of S0s.
Given the smooth transition between `genuine' lenticular and spiral galaxy types, the exact division between the morphological types is difficult to place, leading to the systematic differences in individual visual classifications.
The CNN tends to label such intermediate galaxies as lenticular (e.g. Fig.~\ref{fig:2353168}).
Overall, 50-70 per cent of galaxies classified as S0 by visual inspection were also classified by the CNN as S0 (Fig.~\ref{fig:confusions}). This level of disagreement in the classification of S0 galaxies is relatively similar to that found for observed samples of galaxies \citep[e.g. see figure 14 in][]{Nair_and_Abraham_10}.

\end{itemize}

To summarise, overall the CNN can reasonably accurately distinguish early- and late-type galaxies. Combining the classifications into early- (E+S0) and late-type (Sp+I) galaxies, the CNN is 76-88 per cent accurate over the simulated galaxy sample.
The differences between the CNN and visual classifications (for 4 classes) mainly arise due to differences in how the broad types are subdivided (e.g. late types into Sp and Irr).
For early-type galaxies, the differences are largely due to the difficulty in distinguishing spheroid-dominated (E) and disc-dominated (S0) galaxies from a single image, an issue that also occurs when classifying observed galaxies \citep[c.f.][]{Emsellem_et_al_07, Cappellari_et_al_11}.
For late-type galaxies, the differences appear to be due to how the spiral and irregular classes are defined, in particular whether galaxies with weaker spiral/star-formation features are classified as spiral or irregular.


\bsp	
\label{lastpage}
\end{document}